# ROBUST BAYESIAN COMPRESSIVE SENSING WITH DATA LOSS RECOVERY FOR STRUCTURAL HEALTH MONITORING SIGNALS

Yong Huang[1,2], James L. Beck[2], Stephen Wu[3], Hui Li[1]

[1]Key Lab of Structural Dynamic Behavior and Control of the Ministry of Education,

Harbin Institute of Technology, Harbin 150090, China

[2]Division of Engineering and Applied Science, California Institute of Technology, Pasadena, CA 91125, USA

[3]Computational Science and Engineering Laboratory, Department of Mechanical and Process Engineering,

ETH Zurich, CH-8092 Zurich, Switzerland

**ABSTRACT**

The application of compressive sensing (CS) to structural health monitoring is an emerging research topic. The basic idea in CS is to use a specially-designed wireless sensor to sample signals that are sparse in some basis (e.g. wavelet basis) directly in a compressed form, and then to reconstruct (decompress) these signals accurately using some inversion algorithm after transmission to a central processing unit. However, most signals in structural health monitoring are only approximately sparse, i.e. only a relatively small number of the signal coefficients in some basis are significant, but the other coefficients are usually not exactly zero. In this case, perfect reconstruction from compressed measurements is not expected. A new Bayesian CS algorithm is proposed in which robust treatment of the uncertain parameters is explored, including integration over the prediction-error precision parameter to remove it as a "nuisance" parameter. The performance of the new CS algorithm is investigated using compressed data from accelerometers installed on a space-frame structure and on a cable-stayed bridge. Compared with other state-of-the-art CS methods including our previously-published Bayesian method which uses MAP (maximum a posteriori) estimation of the prediction-error precision parameter, the new algorithm shows superior performance in reconstruction robustness and posterior uncertainty quantification. Furthermore, our method can be utilized for recovery of lost data during wireless transmission, regardless of the level of sparseness in the signal.

**INTRODUCTION**

Structural health monitoring (SHM) is an active and well-established research area that has advanced significantly over the last decade or so. The primary goal is to automatically detect and assess structural damage from severe loading events (e.g. earthquakes, tornados, explosions) or from inevitable aging and degradation, by statistical inference from vibration data (Ou and Li, 2010; Farrar and Worden, 2007; Vanik et al., 2000). The large scales of civil infrastructure systems require sophisticated SHM systems, often with the installation of hundreds of sensors (usually accelerometers). Therefore, data compression techniques are necessary to reduce the cost of transfer and storage for the huge amount of data generated by SHM systems, especially those that are continuously monitoring a large infrastructure system. Furthermore, such techniques provide an effective way to improve the power efficiency and minimize bandwidth during data transmission for wireless monitoring systems (Lynch et al., 2003; Lynch, 2007; Xu et al., 2004). Wavelet-based compression techniques (Xu et al., 2004) and Huffman lossless compression techniques (Lynch et al., 2003) have been developed in recent years. All of these existing data compression methods belong to a conventional framework for sampling signals that follow the Nyquist-Shannon sampling theorem: the sampling rate must be at least twice the maximum frequency present in the signal.

Recently, a new type of sampling theory named *compressive* (or *compressed*) *sensing* (Candes *et al.*, 2006; Donoho, 2006; Candes and Wakin, 2008) has become a very active research topic and shows promise for SHM applications. In a CS sensor, the signal is digitized and linearly projected onto a lower-dimensional space by multiplying with a rectangular matrix. It has been found that a sufficient sparse signal can be reconstructed (decompressed) with high accuracy from a compressed version by using a least-squares method with $l_1$–norm regularization (Candes et al., 2006; Chen et al., 1999), even though the amount of compressed sampling measurements is insufficient by the Nyquist–Shannon criterion. The data reconstruction takes advantage of the signal's sparseness in terms of some basis, allowing only solutions with a small number of nonzero basis coefficients. This new technique presents an efficient signal processing paradigm by merging traditional signal sensing and compression into a single phase, therefore increasing the efficiency of data transfer and storage. Bao et al. (2013) also utilize CS techniques to recover missing data during wireless transmission (Meyer et al., 2010), since data loss is essentially the same as data compression in the CS framework.

In the past few years, many CS reconstruction methods have been proposed to solve the CS reconstruction problem using $l_1$-norm minimization or its extensions (e.g. Candes et al., 2006; Chen et al., 1999; Figueiredo, et al, 2007; Tropp and Gilbert, 2007; Needell and Tropp, 2009). Alternative methods are $l_0$-norm minimization, such as iterative hard-thresholding methods (Blumensath and Davies, 2009; Blumensath, 2012), and non-convex $l_p$ -pseudo-norm ($0 < p < 1$) minimization such as iteratively re-weighted least squares (Chartrand et.al, 2007; Chartrand and Yin, 2008). Recently, Bayesian Compressive Sensing (BCS) (Ji et al., 2008) has been proposed and its robustness has been studied and improved (Huang et al, 2014). Note that all of these reconstruction algorithms are predicated on the prior assumption that the original signal is very sparse in some basis, i.e., only relatively few basis components have nonzero magnitudes, with all remaining components equal to zero.

Civil infrastructure systems are usually subjected to wide-band transient ambient excitation, and SHM signals are also corrupted by a combination of measurement noise and unknown environmental excitation, so they are usually only "approximately sparse", i.e., their main energy (in terms of the sum of the squares of the basis coefficients) is concentrated in only a few basis components, and most of the other components are close to zero, but not exactly zero. The strict *sparseness level* (the total number of the zero components) is therefore low and so highly accurate reconstruction is unlikely to be achieved with limited measurements. The CS reconstruction of approximately sparse signals has attracted interest recently (Barbier et al., 2012; Gilbert et al., 2009; Stojnic et al., 2008).

To adequately treat the uncertainties when establishing a reconstructed signal model from the compressed data, we favor a hierarchical Bayesian approach based on sparse Bayesian learning (SBL) (Tipping, 2001; Tipping and Faul, 2003; Wipf and Rao, 2004; Wipf and Rao, 2006). In contrast to the deterministic CS reconstruction algorithms that provide only a point estimate of the basis coefficients to specify the signal reconstruction, the Bayesian CS algorithm uses posterior probability distributions over these coefficients as an efficient way to quantify uncertainty in the reconstructed signals. The effective dimensionality (number of nonzero signal coefficients) of the signal model is determined automatically as part of the full Bayesian inference procedure, and does not require tuning or knowing the signal sparseness or noise levels that are required in the traditional CS method. Note that many popular compressive sensing recovery schemes can be interpreted using a Bayesian point of view. For example, $l_1$-norm minimization can be formulated as maximum a posteriori (MAP) estimation under a Bayesian model with a sparseness-inducing Laplace prior distribution (Babacan et al., 2010).

In this article, two Bayesian CS algorithms are proposed, one with only MAP estimation of the hyper-parameters, called BCS-MPE, and the second one with integration over the uncertain prediction-error precision parameter, called BCS-IPE. From the presented results with real SHM signals, the superiority of reconstruction robustness and posterior uncertainty quantification of the BCS-IPE algorithm is demonstrated, along with its ability to recover a signal from partial data loss during wireless transmission.

**BAYESIAN COMPRESSIVE SENSING AND PROPOSED ALGORITHMS**

We consider an unknown signal $\bar{\mathbf{x}}$ with $N$ degrees of freedom: $\bar{\mathbf{x}} = [\bar{x}(1), \cdots \bar{x}(N)]^T$ in $\mathbb{R}^N$, which is represented by a set of orthogonal basis vectors as

$$\bar{\mathbf{x}} = \mathbf{\Psi}\bar{\mathbf{w}}_{as} \quad \text{or} \quad \bar{\mathbf{x}} = \mathbf{\Psi}(\bar{\mathbf{w}}_s + \bar{\mathbf{w}}_e) \tag{1}$$

where $\mathbf{\Psi} = [\mathbf{\Psi}_1, \cdots, \mathbf{\Psi}_N]$ is the $N \times N$ orthogonal matrix with the basis of $N \times 1$ vectors $\{\mathbf{\Psi}_n\}_{n=1}^N$ as columns. The unknown vector $\bar{\mathbf{w}}_{as} = \mathbf{\Psi}^T \bar{\mathbf{x}}$ is assumed to be approximately sparse in terms of basis coefficients, and we define $\bar{\mathbf{w}}_s$ to represent an $N$-dimensional $T$-sparse vector that is identical to the vector $\bar{\mathbf{w}}_{as}$ for the $T$ basis components with large magnitude while all other components are zero. The difference between the two vectors $\bar{\mathbf{w}}_{as} - \bar{\mathbf{w}}_s$ is denoted by $\bar{\mathbf{w}}_e$, whose non-zero components are ideally the small magnitude components in $\bar{\mathbf{w}}_{as}$. The total number $(N - T)$ of the zero components of $\bar{\mathbf{w}}_s$ represents the *effective sparseness level* of the signal $\bar{\mathbf{x}}$ with respect to the basis $\{\mathbf{\Psi}_n\}$.

In the CS framework, one infers the signal coefficients vector $\mathbf{w}$ of interest from compressed data instead of directly sampling the actual signal $\bar{\mathbf{x}}$. Let $\mathbf{y}$ in $\mathbb{R}^K$ represent the compressed measurement from $K \ll N$ linear projections of the original signal $\bar{\mathbf{x}}$ using a chosen $K \times N$ random projection matrix $\mathbf{\Phi}$ that is built into the sensor (typically, each element in $\mathbf{\Phi}$ is drawn from zero-mean Gaussian distribution $\mathcal{N}(0, 1)$):

$$\mathbf{y} = \mathbf{\Phi}\bar{\mathbf{x}} + \mathbf{r} \tag{2}$$

where $\mathbf{r}$ represents any measurement noise, which will be relatively small. Incorporating (1), we can rewrite $\mathbf{y}$ as

$$\mathbf{y} = \mathbf{\Theta}\bar{\mathbf{w}}_{as} + \mathbf{r} = \mathbf{\Theta}\bar{\mathbf{w}}_s + \mathbf{\Theta}\bar{\mathbf{w}}_e + \mathbf{r} \tag{3}$$

where $\mathbf{\Theta} = \mathbf{\Phi}\mathbf{\Psi} = [\mathbf{\Theta}_1, \cdots, \mathbf{\Theta}_N]$.

For signal reconstruction, the compressed data $\mathbf{y}$ is represented as:

$$\mathbf{y} = \mathbf{\Theta}\mathbf{w} + \mathbf{\Theta}(\bar{\mathbf{w}}_{as} - \mathbf{w}) + \mathbf{r} = \mathbf{\Theta}\mathbf{w} + \mathbf{\Theta}(\bar{\mathbf{w}}_s + \bar{\mathbf{w}}_e - \mathbf{w}) + \mathbf{r} = \mathbf{\Theta}\mathbf{w} + \mathbf{e} \tag{4}$$

where $\mathbf{e} = \mathbf{\Theta}(\bar{\mathbf{w}}_s + \bar{\mathbf{w}}_e - \mathbf{w}) + \mathbf{r}$ represents the unknown prediction error in $\mathbf{y}$ when the unknown signal is modeled by $\bar{\mathbf{x}} = \mathbf{\Psi}\mathbf{w}$, combined with any measurement noise $\mathbf{r}$. Because we want $\mathbf{w}$ to pick up the large magnitude components of $\bar{\mathbf{w}}_{as}$, we want $\mathbf{e}$ to be small. For data compression, we will have $K \ll N$, so (4) leads to an ill-posed inversion problem to find the sparse weights $\mathbf{w}$, and hence the signal $\mathbf{x}$ in $\mathbb{R}^N$, from data $\mathbf{y}$ in $\mathbb{R}^K$. In order to reduce the number of solutions for such an underdetermined system, one can impose an extra constraint of sparseness by allowing only solutions which have a small number of nonzero basis coefficients. A typical approach is to use an $l_1$–norm regularized formulation to estimate "optimal" basis coefficients:

$$\widetilde{\mathbf{w}} = \arg\ \min\{\|\mathbf{y} - \mathbf{\Theta}\mathbf{w}\|_2^2 + \lambda\|\mathbf{w}\|_1\} \tag{5}$$

where the penalty parameter $\lambda$ controls the trade-off between how well the data is fitted (first term) and how sparse the signal is (second term). Appropriate CS reconstruction algorithms have been proposed, including linear programming (Chen *et al.*, 1999; Candes*et al.*, 2006) and greedy algorithms (Tropp and Gilbert, 2007), based on the framework defined in (5). In contrast, we use sparse Bayesian learning to infer the plausible values of $\mathbf{w}$ based on the compressed data $\mathbf{y}$.

*Sparse Bayesian learning for compressive sensing reconstruction*

In sparse Bayesian learning, Bayes' theorem is applied to find the posterior probability density function (PDF) $p(\mathbf{w}|\mathbf{y})$ for the signal weights $\mathbf{w}$ in (4) based on the linearly projected data $\mathbf{y}$. The uncertain prediction error $\mathbf{e}$ in (4) is modeled as a zero-mean Gaussian vector with unknown $K \times K$ covariance matrix $\beta^{-1}\mathbf{I}_k = \text{diag}(\beta^{-1}, \dots, \beta^{-1})$. This maximum entropy probability model gives the largest uncertainty for $\mathbf{e}$ subject to the first two moment constraints: $\mathbf{E}[e_k] = 0, \mathbf{E}[e_k^2] = \beta^{-1}, k = 1, \dots, K$. Thus, one gets a Gaussian likelihood function for parameters $\mathbf{w}$ and $\beta$ based on the observed CS measurement $\mathbf{y}$:

$$p(\mathbf{y}|\mathbf{w},\beta) = (2\pi\beta^{-1})^{-\frac{K}{2}}\exp\left(-\frac{\beta}{2}\|\mathbf{y} - \mathbf{\Theta}\mathbf{w}\|_2^2\right) \tag{6}$$

The logarithm of this likelihood corresponds to the first term of (5) in the deterministic CS data inversion. A Gamma conjugate prior is taken for $\beta$:

$$p(\beta|a_0, b_0) = \text{Gamma}(\beta|a_0, b_0) = \frac{b_0^{a_0}}{\Gamma(a_0)}\beta^{a_0-1}\exp(-b_0\beta) \tag{7}$$

In the Bayesian formulation, the constraint of sparseness is formalized by placing a sparseness-promoting prior on the signal basis coefficients $\mathbf{w}$. In the sparse Bayesian learning approach, a special Gaussian prior distribution is used which is known as the *automatic relevance determination prior* (ARD prior):

$$p(\mathbf{w}|\mathbf{\Sigma}_0) = \prod_{n=1}^{N} N(w_n|0, \sigma_n^2) = \prod_{n=1}^{N}\left[(2\pi\sigma_n^2)^{-1/2}\exp\left\{-\frac{1}{2}\sigma_n^{-2}w_n^2\right\}\right] \tag{8}$$

where the prior covariance matrix $\mathbf{\Sigma}_0 = \text{diag}(\sigma_1^2, \dots, \sigma_N^2)$ and $\sigma_n^2$ is the prior variance for $w_n$. Tipping (2001) has shown that maximizing the model evidence (see below for its definition) with respect to all of the $\sigma_n^2$ controls the model sparseness because many $\sigma_n^2 \to 0$, implying $w_n \to 0$, thereby having an effect similar to the regularization term in (5). In order to allow the tractability of an integration involved later, we incorporate the prediction error precision $\beta$ in the ARD prior by replacing each $\sigma_n^2$ by $\alpha_n = 1/(\beta\sigma_n^2)$ to get the following prior PDFs:

$$p(\mathbf{w}|\boldsymbol{\alpha},\beta) = p(\mathbf{w}|\boldsymbol{\Sigma}_0) = \prod_{n=1}^{N}\mathcal{N}(w_n|0,\beta^{-1}\alpha_n^{-1}) = \prod_{n=1}^{N}\left[(2\pi)^{-1/2}(\beta\alpha_n)^{1/2}\exp\left\{-\frac{1}{2}\beta\alpha_n w_n^2\right\}\right] \qquad (9)$$

The likelihood function in (6) for the CS measurements $\mathbf{y}$ and the prior on $\mathbf{w}$ in (9) define a *stochastic model class* $\mathcal{M}(\boldsymbol{\alpha},\beta)$ for the signal model, which has the posterior distribution $p(\mathbf{w}|\mathbf{y},\boldsymbol{\alpha},\beta)$ over the basis coefficients given by Bayes' theorem:

$$p(\mathbf{w}|\mathbf{y},\boldsymbol{\alpha},\beta) = p(\mathbf{y}|\mathbf{w},\beta)p(\mathbf{w}|\boldsymbol{\alpha},\beta)/p(\mathbf{y}|\boldsymbol{\alpha},\beta) \qquad (10)$$

where the normalizing constant for the posterior PDF is the *evidence* (or *marginal likelihood*) for $\mathcal{M}(\boldsymbol{\alpha},\beta)$:

$$p(\mathbf{y}|\boldsymbol{\alpha},\beta) = \int p(\mathbf{y}|\mathbf{w},\beta)p(\mathbf{w}|\boldsymbol{\alpha},\beta)\mathbf{dw} = \mathcal{N}(\mathbf{y}|0,\beta^{-1}\mathbf{B}) \qquad (11)$$

where $\qquad \mathbf{B} = \mathbf{I}_K + \boldsymbol{\Theta}\mathbf{A}^{-1}\boldsymbol{\Theta}^T, \quad \mathbf{A} = \mathrm{diag}(\alpha_1,\dots,\alpha_N).$

It is readily shown that the posterior PDF is a multivariate Gaussian distribution:

$$p(\mathbf{w}|\mathbf{y},\boldsymbol{\alpha},\beta) = \mathcal{N}(\mathbf{w}|\boldsymbol{\mu},\boldsymbol{\Sigma}) \qquad (12)$$

with mean and covariance matrix:

$$\boldsymbol{\mu} = \mathbf{C}^{-1}\boldsymbol{\Theta}^T\mathbf{y}, \qquad (13a)$$

$$\boldsymbol{\Sigma} = \beta^{-1}\mathbf{C}^{-1}, \qquad (13b)$$

where $\qquad \mathbf{C} = \mathbf{A} + \boldsymbol{\Theta}^T\boldsymbol{\Theta}. \qquad (13c)$

For given hyper-parameters $\boldsymbol{\alpha}$ and $\beta$ defining the model class $\mathcal{M}(\boldsymbol{\alpha},\beta)$, the posterior probability distribution for the reconstructed signal $\mathbf{x} = \boldsymbol{\Psi}\mathbf{w}$ is Gaussian, $p(\mathbf{x}|\mathbf{y},\boldsymbol{\alpha},\beta) = \mathcal{N}(\mathbf{x}|\boldsymbol{\mu}_\mathbf{x},\boldsymbol{\Sigma}_\mathbf{x})$ with mean and covariance matrix:

$$\boldsymbol{\mu}_\mathbf{x} = \boldsymbol{\Psi}\boldsymbol{\mu} = \boldsymbol{\Psi}\mathbf{C}^{-1}\boldsymbol{\Theta}^T\mathbf{y} \qquad (14a)$$

$$\boldsymbol{\Sigma}_\mathbf{x} = \boldsymbol{\Psi}\boldsymbol{\Sigma}\boldsymbol{\Psi}^T = \beta^{-1}\boldsymbol{\Psi}\mathbf{C}^{-1}\boldsymbol{\Psi}^T \qquad (14b)$$

If the model class is globally identifiable, so that the posterior PDF $p(\boldsymbol{\alpha},\beta|\mathbf{y})$ has a single pronounced global maximum with respect to $\boldsymbol{\alpha}$ and $\beta$, then the posterior PDF $p(\mathbf{w}|\mathbf{y})$ can be estimated accurately using the most probable model class $\mathcal{M}(\hat{\boldsymbol{\alpha}},\hat{\beta})$ by Laplace's asymptotic approximation (Beck and Katafygiotis, 1998; Beck, 2010):

$$p(\mathbf{w}|\mathbf{y}) = \int p(\mathbf{w}|\mathbf{y}, \boldsymbol{\alpha}, \beta) p(\boldsymbol{\alpha}, \beta|\mathbf{y}) d\boldsymbol{\alpha} d\beta$$

$$\approx p(\mathbf{w}|\mathbf{y}, \hat{\boldsymbol{\alpha}}, \hat{\beta}) \qquad (15)$$

where $(\hat{\boldsymbol{\alpha}}, \hat{\beta}) = \arg\max_{[\boldsymbol{\alpha}, \beta]} p(\boldsymbol{\alpha}, \beta|\mathbf{y})$.

We note that this maximization is not convex with respect to $\boldsymbol{\alpha}$ and there are local maxima that may trap an optimization algorithm as $K$ becomes much smaller than $N$, producing non-robust signal reconstruction, as noted in Huang et al. (2014).

*Algorithm with MAP estimation of hyper-parameters $\beta$ and $\boldsymbol{\alpha}$*

We first develop an algorithm that is based on the MAP values of all of the uncertain hyper-parameters: $\boldsymbol{\alpha}, \beta, a_0$ and $b_0$, which are modeled as mutually independent a priori. By taking a uniform prior on $\boldsymbol{\alpha}, a_0$ and $b_0$ and Gamma prior on $\beta$ as in (7), and utilizing (11), we need to maximize the posterior PDF:

$$p(\boldsymbol{\alpha}, \beta, a_0, b_0|\mathbf{y}) \propto p(\mathbf{y}|\boldsymbol{\alpha}, \beta) p(\beta|a_0, b_0)$$

$$= \mathcal{N}(\mathbf{y}|0, \beta^{-1}\mathbf{B}) \text{Gamma}(\beta|a_0, b_0)$$

$$= (2\pi)^{-\frac{K}{2}} \frac{b_0^{a_0}}{\Gamma(a_0)} \beta^{a_0-1+K/2} (|\mathbf{B}|)^{-\frac{1}{2}} \exp\left\{-b_0\beta - \frac{1}{2}\beta \mathbf{y}^T \mathbf{B}^{-1} \mathbf{y}\right\} \qquad (16)$$

The optimal values of the parameters, $\hat{\beta}, \hat{a}_0$ and $\hat{b}_0$ satisfy the following equations obtained by taking the derivatives of the logarithm of (16) with respect to $\beta$, $a_0$ and $b_0$, respectively:

$$\hat{\beta} = \frac{K + 2(\hat{a}_0 - 1)}{\mathbf{y}^T \mathbf{B}^{-1} \mathbf{y} + 2\hat{b}_0} \qquad (17)$$

$$\log \hat{a}_0 - \psi(\hat{a}_0) = 0 \qquad (18)$$

$$\hat{b}_0 = \hat{a}_0 \hat{\beta}^{-1} \qquad (19)$$

where $\psi(a_0)$ is the Digamma function. Optimization of parameter $a_0$ from (18) leads to $\hat{a}_0 \to \infty$. However, if we substitute (19) into (17), the optimal value $\hat{\beta}$ is given by:

$$\hat{\beta} = \frac{K - 2}{\mathbf{y}^T \mathbf{B}^{-1} \mathbf{y}} \qquad (20)$$

Therefore, the actual optimal values $\hat{a}_0$ and $\hat{b}_0$ are not needed to find the MAP values of $\boldsymbol{\alpha}$ and $\beta$.

Since $p(\boldsymbol{\alpha})$ is constant over the important region of the $\boldsymbol{\alpha}$ space, the optimization over $\boldsymbol{\alpha}$ of the posterior in (16) given $[\beta, a_0, b_0]$, is equivalent to maximizing the log of the evidence function in (11): $\mathcal{L}(\boldsymbol{\alpha}, \beta) = \log p(\mathbf{y}|\boldsymbol{\alpha}, \beta) = \log \mathcal{N}(\mathbf{y}|0, \beta^{-1}\mathbf{B})$. In the original sparse Bayesian learning method of Tipping (2001), the

maximum is found by iterative solution of the stationarity equations obtained from direct differentiation of the log evidence function $\mathcal{L}(\boldsymbol{\alpha}, \beta)$ with respect to $\boldsymbol{\alpha}$. The drawback is that the optimization involves inversion of matrices of size $N \times N$ in the beginning iterations (although for compressive sensing decompression where $K \ll N$, we can use the Woodbury inversion identity to reduce the algorithm to an inversion of a matrix with $\mathbf{O}(K^3)$ multiplications), thereby making this approach relatively slow for reconstruction of CS signals with large dimensions. In practice, the sparse signal models that are finally reconstructed have far fewer non-zero terms than $K$ or $N$, and so the strategy of optimization for a full signal model in the beginning of the iterations seems wasteful. We call this strategy the Top-down SBL algorithm.

There is a faster strategy, which we call the Bottom-up SBL algorithm, that starts with no terms in the basis expansion and then adds relevant ones to the signal model as the iterations proceed; this is done by updating a single hyper-parameter $\alpha_n$ at each iteration to monotonically increase the evidence (Tipping and Faul, 2003). The algorithm is derived by isolating the contribution of each single hyper-parameter $\alpha_n$ in the log evidence function $\mathcal{L}(\boldsymbol{\alpha}, \beta)$ in a convenient form: $\mathcal{L}(\boldsymbol{\alpha}, \beta) = \mathcal{L}(\boldsymbol{\alpha}_{-n}, \beta) + l(\alpha_n, \beta)$, where $\mathcal{L}(\boldsymbol{\alpha}_{-n}, \beta)$ is the log evidence with the component $\alpha_n$ removed. Setting the derivative of $l(\alpha_n, \beta)$ with respect to $\alpha_n$ equal to zero leads to:

$$\hat{\alpha}_n = \begin{cases} \infty, & \text{if } \hat{\gamma}_n \leq 0 \\ \hat{\gamma}_n, & \text{if } \hat{\gamma}_n > 0 \end{cases} \quad (21)$$

where
$$\hat{\gamma}_n = \frac{s_n^2}{\beta q_n^2 - s_n} \quad (22)$$

and the 'sparseness factor' $s_n$ and 'quality factor' $q_n$ are defined by:

$$s_n = \boldsymbol{\Theta}_n^T \mathbf{B}_{-n}^{-1} \boldsymbol{\Theta}_n \quad (23)$$

$$q_n = \boldsymbol{\Theta}_n^T \mathbf{B}_{-n}^{-1} \mathbf{y} \quad (24)$$

where $\mathbf{B}_{-n}$ is $\mathbf{B}$ with the contribution of basis vector $\boldsymbol{\Theta}_n$ removed (Tipping and Faul, 2003). The calculation of $s_n$ and $q_n$ only requires the numerical inversion of an $N' \times N'$ matrix, where $N'$ is the number of non-zero terms in the current signal model and it is much smaller than the number of compressed measurements $K$ in this strategy. Finally, only the components that have finite $\alpha_n$ are retained in the signal model since each $w_n$ with $\alpha_n = \infty$ has prior mean and variance both zero, giving $w_n = 0$, and so its term drops out of (4).

Notice that the optimal $\hat{\beta}$ in (20) depends on $\boldsymbol{\alpha}$ through $\mathbf{B}$ and the optimal $\hat{\boldsymbol{\alpha}}$ from (21) depends on $\beta$ through $\hat{\gamma}_n$. Therefore, an iterative scheme is required for the full optimization of the evidence with respect to $[\boldsymbol{\alpha}, \beta]$. We have found that some care is needed because of the important influence of $\beta$ on the evidence function but that successive relaxation works well (Huang et al., 2014): first $\boldsymbol{\alpha}$ is optimized with $\beta$ fixed and then $\beta$ is optimized with $\alpha$ fixed at its intermediate optimal value, with this procedure being repeated until convergence is achieved. To initialize the algorithm, we select the single $\alpha_n$ of the $n^{th}$ term that maximizes $\|\Theta_n^T \mathbf{y}\|^2/\|\Theta_n\|^2$ over $n$ as $\alpha_n = 1$. All other $\alpha_n's$ are set to infinity, implying that the corresponding terms are excluded in the initial signal model. We update the prediction error precision $\beta$ based on this signal model, then fix this updated $\beta$ and optimize the intermediate evidence function to obtain a set of improved optimal $\alpha_n's$.

The idea of successive relaxation produces a Bayesian CS reconstruction method that iterates between inner and outer loop optimizations. The outer loop updates the prediction-error precision $\beta$ and is terminated when the changes in the reconstructed signal models are sufficiently small, e.g. $\|(\hat{\mathbf{x}})^{[j+1]} - (\hat{\mathbf{x}})^{[j]}\|_2^2 / \|(\hat{\mathbf{x}})^{[j]}\|_2^2 < \epsilon$, a specified threshold (we take $\epsilon = 0.1$ in the examples with real SHM signals later). The inner loop performs the optimization procedure over the prior variances $\alpha$ and is terminated when the change in all $\log \alpha_n's$ is less than $10^{-6}$. We call this procedure Algorithm BCS-MPE, where 'MPE' denotes MAP estimation of the prediction-error parameter $\beta$.

In the inner loop, adding, deleting or re-estimating a basis vector $\Theta_n$ in each iteration is based on whatever gives the maximum log evidence increase $\Delta \mathcal{L}_n$ (Tipping and Faul, 2003). To make the comparison, we need to calculate the increase of log evidence $\Delta \mathcal{L}_n$ for each term in the inner loop. Based on the isolated contribution of the term in the evidence function for each term, it is estimated as $\Delta \mathcal{L}_n = l(\hat{\alpha}_n, \hat{\beta}) - l(\alpha_n, \hat{\beta})$, where $\alpha_n$ is the current hyper-parameter value before optimization, and $\hat{\alpha}_n$ and $\hat{\beta}$ are the optimal hyper-parameters for the prediction error from the previous outer loop (see Step 2 or 12 below). More efficient updating formulae with reduced computation that avoid any matrix inversions are given in Appendix A. The posterior mean $\mu$ and covariance matrix $\Sigma$ contain only those $N' < N$ basis terms that are currently included in the signal model, and the computation thus involves only a small fraction of the full set of basis coefficients.

---

**Algorithm BCS-MPE**

1. Inputs: $\Theta$, $\mathbf{y}$; Outputs: posterior mean and covariance of $\mathbf{w}$ and $\mathbf{x}$

2. Initialize all $\alpha_n's$ to a very large value except set $\alpha_n = 1$ for the term that maximizes $\|\Theta_n^T \mathbf{y}\|^2/\|\Theta_n\|^2$, then initialize $\beta$ based on the current $\alpha_n$ using (20).

3. Compute initial $\mu$ and $\Sigma$ using (13) (both are scalars because $N' = 1$ initially), and

   initialize $s_m$ and $q_m$ for all terms $(m = 1, ..., N)$, using the formulae in Appendix A.

4. **While** convergence criterion on the $\log \alpha_n$'s is not met (Inner loop)

5. Select the basis vector $\mathbf{\Theta}_n$ with the largest log evidence increase $\Delta \mathcal{L}_n$

6. **If** $\beta q_n^2 - s_n > 0$ and $\alpha_n = \infty$, add $\mathbf{\Theta}_n$ and update $\alpha_n$ using (21)

7. **If** $\beta q_n^2 - s_n > 0$ and $\alpha_n < \infty$, re-estimate $\alpha_n$ using (21)

8. **If** $\beta q_n^2 - s_n \leq 0$ and $\alpha_n < \infty$, delete $\mathbf{\Theta}_n$ and set $\alpha_n = \infty$

9. **End if**

10. Update $\boldsymbol{\mu}$ and $\boldsymbol{\Sigma}$, and $s_m$ and $q_m$ for all terms $(m = 1, ..., N)$, using the formulae in Appendix A.

11. **End while** (the intermediate optimal $\alpha_n$'s are then established for the current $\beta$)

12. Update $\beta$ based on the current $\alpha_n$'s using (20).

13. Check if updating of mean reconstructed signal model $\hat{\mathbf{x}} = \boldsymbol{\Psi}\boldsymbol{\mu}$, has converged. If not, use the current intermediate optimal hyper-parameters ($\boldsymbol{\alpha}$ and $\beta$) as initial values and repeat the inner loop (steps 3 to 10); otherwise end.

  Note that maximizing the log evidence function $\mathcal{L}(\boldsymbol{\alpha}, \beta)$ can achieve an optimal balance between data fitting and signal model sparseness automatically and it has an interesting information-theoretic interpretation (Beck, 2010; Huang *et al*, 2014). It is shown that the log evidence can be represented as the difference between the average data-fit term and model complexity (more sparseness corresponds to less model complexity) of the signal model $\mathcal{M}(\boldsymbol{\alpha}, \beta)$. In Algorithm BCS-MPE, the MAP values of hyper-parameters $\boldsymbol{\alpha}$ and $\beta$ are used in (15) to approximate $p(\mathbf{w}|\mathbf{y})$ under the assumption that $p(\boldsymbol{\alpha}, \beta|\mathbf{y})$ is sharply peaked around its mode. This is an effective assumption for CS reconstruction in the high sparseness case, because there is have a high chance to find a sufficiently sparse signal model given by the posterior mean coefficients $\mathbf{E}(\mathbf{w}|\mathbf{y}, \hat{\boldsymbol{\alpha}}, \hat{\beta})$ that can fit the measurement vector $\mathbf{y}$ well, so the algorithmic optimization is likely to give the MAP values $(\hat{\boldsymbol{\alpha}}, \hat{\beta})$. However, MAP estimation has the drawback that the uncertainty quantified by the full posterior PDF $p(\mathbf{w}|\mathbf{y})$ is underestimated when the other plausible models specified by $\boldsymbol{\alpha}, \beta$ are ignored.

  It has also been found that current BCS algorithms that are based on MAP estimation of all of the hyper-parameters suffer from a robustness problem (Huang *et al.*, 2014): when the number of compressed measurements $K$ is a lot less than the number of signal degrees of freedom $N$, sometimes the algorithm finds only local maxima of the evidence that correspond to larger amounts of non-zero signal components and large reconstruction errors. For these suboptimal models $(\breve{\boldsymbol{\alpha}}, \breve{\beta})$, the Laplace asymptotic approximation is very poor because the dominant contribution to the integral in Eq. (15) is missed.

Another source of poor robustness is that real structural health monitoring signals are usually only approximately sparse. In this case, the signal model for a relatively sparse vector $\bar{\mathbf{w}}_s$ in (3) cannot fit the measurement vector $\mathbf{y}$ well, because of the larger error $\boldsymbol{\Theta}\bar{\mathbf{w}}_e + \mathbf{r}$, while $\bar{\mathbf{w}}_{as}$ in (3) can fit $\mathbf{y}$ with much smaller error $\mathbf{r}$ but the signal model will no longer be sparse. Although this trade-off between sparseness and data-fitting is always present, in the case of approximately sparse data $\mathbf{y}$, it will cause a spreading out of the global peak of the evidence function $p(\mathbf{y}|\boldsymbol{\alpha}, \beta)$ around the optimal values $(\widehat{\boldsymbol{\alpha}}, \widehat{\beta})$ and this peak will not be sharp. Using just the MAP estimates of $(\boldsymbol{\alpha}, \beta)$ that maximize the evidence function $p(\mathbf{y}|\boldsymbol{\alpha}, \beta)$ may not lead to a robust reconstruction because it ignores the other values of $(\boldsymbol{\alpha}, \beta)$ that contribute significantly to the integral in Eq. (15) (that is, their evidence values are close to the maximum value occurring at $(\widehat{\boldsymbol{\alpha}}, \widehat{\beta})$).

The robust way to tackle this problem is to account for the full posterior uncertainty of the hyper-parameters $(\boldsymbol{\alpha}, \beta)$ by integrating them out as 'nuisance' parameters, that is, by marginalizing them. Although it turns out that marginalizing $\boldsymbol{\alpha} \in \mathbb{R}^N$ is analytically intractable, the prior $p(\mathbf{w}, \beta|\boldsymbol{\alpha}, a_0, b_0) = p(\mathbf{w}|\boldsymbol{\alpha}, \beta)p(\beta|a_0, b_0)$ given by Eqs. (9) and (7) is conjugate to the likelihood $p(\mathbf{y}|\mathbf{w}, \beta)$ given by Eq. (6) and so both $\mathbf{w}$ and $\beta$ can be integrated out analytically, which can improve robustness during the optimization of $\boldsymbol{\alpha}$.

In the next section, a sparse Bayesian learning algorithm is proposed that integrates out the uncertain prediction-error precision parameter $\beta$ to make the BCS decompression algorithm more robust. We believe this algorithm is a new contribution. Ji et al. (2009) integrated out the prior uncertainty on $\beta$ but not the posterior uncertainty to get the marginal posterior for $\mathbf{w}$.

*Algorithm with marginalization of prediction error $\beta$ and MAP estimation of $\boldsymbol{\alpha}$*

The posterior PDF for $\beta$ is readily obtained by Bayes' theorem using the likelihood in (11) and prior in (7):

$$p(\beta|\mathbf{y}, \boldsymbol{\alpha}, a_0, b_0) \propto p(\mathbf{y}|\boldsymbol{\alpha}, \beta)p(\beta|a_0, b_0)$$

$$= \frac{\beta^{a_0 - 1 + \frac{K}{2}}}{(2\pi)^{\frac{K}{2}}} \frac{b_0^{a_0}}{\Gamma(a_0)} |\mathbf{B}|^{-\frac{1}{2}} \exp\left\{-\frac{\beta}{2}\mathbf{y}^T \mathbf{B}^{-1}\mathbf{y} - b_0 \beta\right\} \propto \text{Gamma}(\beta|a_0', b_0'). \tag{25}$$

From observation of Eq. (25), it is obvious that the posterior PDF of $\beta$ is a Gamma distribution Gamma$(\beta|a_0', b_0')$ where the shape and rate parameters become:

$$a_0' = a_0 + K/2 \tag{26}$$

$$b_0' = b_0 + \mathbf{y}^T \mathbf{B}^{-1} \mathbf{y}/2. \tag{27}$$

According to the Total Probability Theorem, we get for the posterior $p(\mathbf{w}|\mathbf{y}, \boldsymbol{\alpha}, a_0, b_0)$:

$$p(\mathbf{w}|\mathbf{y}, \boldsymbol{\alpha}, a_0, b_0) = \int p(\mathbf{w}|\mathbf{y}, \boldsymbol{\alpha}, \beta) p(\beta|\mathbf{y}, \boldsymbol{\alpha}, a_0, b_0) d\beta$$

$$= \frac{\Gamma(a_0' + K/2)(a_0'/b_0')^{1/2}}{\Gamma(a_0')(2\pi a_0')^{K/2}|\mathbf{C}|^{-1/2}} \left\{ 1 + \frac{1}{2a_0'} \left[ \frac{a_0'}{b_0'} (\mathbf{w} - \mathbf{C}^{-1}\boldsymbol{\Theta}^T\mathbf{y})^T \mathbf{C} (\mathbf{w} - \mathbf{C}^{-1}\boldsymbol{\Theta}^T\mathbf{y}) \right] \right\}^{-K/2}$$

$$= \text{St}\left(\mathbf{w}|\boldsymbol{\mu}, \frac{a_0'}{b_0'}\mathbf{C}, 2a_0'\right) \tag{28}$$

where $\boldsymbol{\mu}$ and $\mathbf{C}$ are given by (13a) and (13c). This multivariate form of Student's t-distribution has mean and covariance matrix:

$$\mathbf{E}(\mathbf{w}|\mathbf{y}, \boldsymbol{\alpha}, a_0, b_0) = \boldsymbol{\mu} = \mathbf{C}^{-1}\boldsymbol{\Theta}^T\mathbf{y} = \boldsymbol{\Lambda}\boldsymbol{\Theta}^T\mathbf{y} \tag{29a}$$

$$\mathbf{Cov}(\mathbf{w}|\mathbf{y}, \boldsymbol{\alpha}, a_0, b_0) = \frac{b_0'}{a_0'-1}\mathbf{C}^{-1} = \hat{\beta}(a_0, b_0)^{-1}\boldsymbol{\Lambda} \tag{29b}$$

where $\boldsymbol{\Lambda} = \mathbf{C}^{-1}$ and $\hat{\beta}(a_0, b_0)$ is given by (17), which agree with the mean $\boldsymbol{\mu}$ and covariance matrix $\boldsymbol{\Sigma}$ given in (13a), (13b) for BCS-MPE before utilizing (19) in (17). It follows from $\mathbf{x} = \boldsymbol{\Psi}\mathbf{w}$ that for given hyper-parameters $\boldsymbol{\alpha}, a_0$ and $b_0$ defining the stochastic model class $\mathcal{M}(\boldsymbol{\alpha}, a_0, b_0)$, the posterior probability distribution for the reconstructed signal $\mathbf{x}$ is a Student's t-distribution with mean and covariance matrix corresponding to (14a) and (14b):

$$\mathbf{E}(\mathbf{x}|\mathbf{y}, \boldsymbol{\alpha}, a_0, b_0) = \boldsymbol{\Psi}\boldsymbol{\mu} = \boldsymbol{\Psi}\mathbf{C}^{-1}\boldsymbol{\Theta}^T\mathbf{y} = \boldsymbol{\Psi}\boldsymbol{\Lambda}\boldsymbol{\Theta}^T\mathbf{y} \tag{30a}$$

$$\mathbf{Cov}(\mathbf{x}|\mathbf{y}, \boldsymbol{\alpha}, a_0, b_0) = \frac{b_0'}{a_0'-1}\boldsymbol{\Psi}\mathbf{C}^{-1}\boldsymbol{\Psi}^T = \hat{\beta}(a_0, b_0)^{-1}\boldsymbol{\Psi}\boldsymbol{\Lambda}\boldsymbol{\Psi}^T \tag{30b}$$

To find the MAP values of the hyper-parameters $[\boldsymbol{\alpha}, a_0, b_0]$, we need the corresponding evidence function, which is given by:

$$p(\mathbf{y}|\boldsymbol{\alpha}, a_0, b_0) = \int p(\mathbf{y}|\boldsymbol{\alpha}, \beta) p(\beta|a_0, b_0) d\beta$$

$$= \frac{\Gamma(a_0 + K/2)(a_0/b_0)^{1/2}}{\Gamma(a_0)(2\pi a_0)^{K/2}|\mathbf{B}|^{1/2}} \left\{ 1 + \frac{1}{2a_0}\left(\frac{a_0}{b_0}\mathbf{y}^T\mathbf{B}^{-1}\mathbf{y}\right) \right\}^{-K/2 - a_0}$$

$$= \text{St}\left(\mathbf{y}|\mathbf{0}, \frac{a_0}{b_0}\mathbf{B}^{-1}, 2a_0\right) \tag{31}$$

Since $p(\boldsymbol{\alpha}), p(a_0)$ and $p(b_0)$ are taken as constant over the important regions of their parameter spaces, the optimization over $\boldsymbol{\alpha}, a_0$ and $b_0$ to find their MAP values is equivalent to maximizing the log evidence function $\mathcal{L}(\boldsymbol{\alpha}, a_0, b_0) = \log p(\mathbf{y}|\boldsymbol{\alpha}, a_0, b_0)$. By solving the stationarity equations obtained from direct differentiation of $\mathcal{L}(\boldsymbol{\alpha}, a_0, b_0)$ with respect to $a_0$ and $b_0$, the optimal estimates $\tilde{a}_0$ and $\tilde{b}_0$ are given by:

$$\tilde{b}_0 = \frac{\tilde{a}_0}{K} \mathbf{y}^{\text{T}} \mathbf{B}^{-1} \mathbf{y} \tag{32}$$

$$\psi(\tilde{a}_0 + K/2) - \psi(\tilde{a}_0) - \log(1 + \mathbf{y}^{\text{T}} \mathbf{B}^{-1} \mathbf{y}/2\tilde{b}_0) = 0 \tag{33}$$

The solution of these equations leads to $\tilde{a}_0 \to \infty$. However, from (26), we see that the effect of $K$ measurement points is to increase the value of the coefficient $a_0'$ by $K/2$. Thus we can consider the parameter $a_0$ in the prior in terms of $2a_0$ 'effective' prior measurements. Since our current knowledge of the true prediction error is little, the best choice is to make the shape parameter $a_0$ much smaller than $K/2$, so that the prior distribution of $\beta$ has little effect on its posterior distribution. If we impose only the constraint of mean $\mathbf{E}(\beta|b_0) = 1/b_0$, the maximum entropy prior for $\beta$ is the exponential distribution, which is a special case of the Gamma distribution with shape parameter $a_0 = 1$. Since the maximum entropy prior PDF gives the largest uncertainty for $\beta$, subject to only the mean constraint, we choose it, and so set $a_0 = 1$ and then use (32) for the optimal value $\tilde{b}_0$.

To analyze the dependence of $\mathcal{L}(\boldsymbol{\alpha})$ on a single hyper-parameter $\alpha_n$, we rewrite the log evidence $\mathcal{L}(\boldsymbol{\alpha}) = \mathcal{L}(\boldsymbol{\alpha}_{-n}) + l(\alpha_n)$. where $\mathcal{L}(\boldsymbol{\alpha}_{-n})$ is the log evidence with the component $\alpha_n$ removed (Tipping and Faul, 2003). Setting the derivative of $l(\alpha_n)$ with respect to $\alpha_n$ equal to zero leads to the optimal value:

$$\tilde{\alpha}_n = \begin{cases} \infty, & \text{if } \tilde{\gamma}_n \leq 0 \\ \tilde{\gamma}_n & \text{if } \tilde{\gamma}_n > 0 \end{cases} \tag{34}$$

where 
$$\tilde{\gamma}_n = \frac{s_n^2 - s_n q_n^2/g_n}{(K+2a_0)q_n^2/g_n - s_n}. \tag{35}$$

and 
$$g_n = \mathbf{y}^T \mathbf{B}_{-n}^{-1} \mathbf{y} + 2b_0. \tag{36}$$

The 'sparseness factor' $s_n$ and 'quality factor' $q_n$ are given by (23) and (24). Similar to the algorithm BCS-MPE, we employ successive relaxation to optimize the evidence in an iterative fashion: first, $\boldsymbol{\alpha}$ is optimized with $b_0$ fixed and then $b_0$ is updated using (32) with $\boldsymbol{\alpha}$ fixed at its intermediate optimal value. This procedure is repeated until convergence is achieved. The procedure is summarized below in Algorithm BCS-IPE, where 'IPE' denotes integration over the prediction-error precision parameter $\beta$. The termination criteria for the inner and outer loops are the same as for BCS-MPE. The updating formulae in Appendix B are used to give a more efficient implementation that does not require any matrix inversions.

**Algorithm BCS-IPE**

1. Inputs: $\boldsymbol{\Theta}$, $\mathbf{y}$; Outputs: posterior mean and covariance of $\mathbf{w}$ and $\mathbf{x}$

2. Initialize all $\alpha_n$'s to a very large value except set $\alpha_n = 1$ for the term that maximizes $\|\boldsymbol{\Theta}_n^T\mathbf{y}\|^2/\|\boldsymbol{\Theta}_n\|^2$. Set parameter $b_0 = 0$.

3. Compute initial $\boldsymbol{\mu}$ and $\boldsymbol{\Lambda} = \mathbf{C}^{-1}$ using (13a,c) (both are scalars because $N' = 1$ initially), and initialize $s_m$, $q_m$ and $g_m$ for all terms $(m = 1, \ldots, N)$, using the formulae in Appendix B.

4. **While** convergence criterion on the $\log \alpha_n$'s is not met (Inner loop)

5. Select the basis vector $\boldsymbol{\Theta}_n$ with the largest log evidence increase $\Delta \mathcal{L}_n$

6. **If** $\tilde{\gamma}_n > 0$ and $\alpha_n = \infty$, add $\boldsymbol{\Theta}_n$ and update $\alpha_n$ using (34)

7. **If** $\tilde{\gamma}_n > 0$ and $\alpha_n < \infty$, re-estimate $\alpha_n$ using (34)

8. **If** $\tilde{\gamma}_n \leq 0$ and $\alpha_n < \infty$, delete $\boldsymbol{\Theta}_n$ and set $\alpha_n = \infty$

9. **End if**

10. Update $\boldsymbol{\mu}$ and $\boldsymbol{\Lambda}$, and $s_m$, $q_m$ and $g_m$ for all terms $(m = 1, \ldots, N)$, using the formulae in Appendix B.

11. **End while** (the intermediate optimal $\alpha_n$'s are then established for the current $b_0$)

12. Update $b_0$ using (32) based on the current $\alpha_n$'s, i.e. $b_0 = \frac{1}{K}\mathbf{y}^T\mathbf{B}^{-1}\mathbf{y}$.

13. Check if updating of mean reconstructed signal model $\hat{\mathbf{x}} = \boldsymbol{\Psi}\boldsymbol{\mu}$, has converged. If not, use the current intermediate optimal hyper-parameters ($\boldsymbol{\alpha}$ and $b_0$) as initial values and repeat the inner loop (steps 3 to 10); otherwise end.

Ji et al. (2009) also marginalize the prediction-error parameter $\beta$ out. However, in their examples, they fix the shape parameter $a_0$ and rate parameter $b_0$ of the Gamma prior PDF over $\beta$ as the values $a_0 = (10/std(\mathbf{y}))^2$ and $b_0 = 1$, while our algorithm has $a_0 = 1$ and optimizes over $b_0$ and $\boldsymbol{\alpha}$ by employing successive relaxation to maximize the evidence for the model class $\mathcal{M}(\boldsymbol{\alpha}, b_0)$, thereby updating all hyper-parameters effectively. In addition, for the marginal posterior on $\mathbf{w}$ in (28), we integrate out the posterior uncertainty on $\mathbf{w}$ whereas Ji et al. (2009) incorrectly integrates out the prior uncertainty on $\beta$. As a result, their approach to integrating $\beta$ does not lead to much improvement (e.g. Fig. 4 in Ji et al. (2009)). In contrast, we show later that our proposed BCS-IPE algorithm leads to much better CS reconstructions of approximately sparse SHM signals than the BCS-MPE algorithm. Furthermore, because BCS-IPE uses the posterior PDF $p(\mathbf{w}|\mathbf{y},\widetilde{\boldsymbol{\alpha}})$, it gives more reliable posterior uncertainty quantification for $\mathbf{w}$ than BCS-MPE. It therefore provides a powerful diagnostic tool for whether the decompressed representation $\mathbf{E}(\mathbf{x}|\mathbf{y},\widetilde{\boldsymbol{\alpha}})$ of the signal is accurate or not.

*Comparison between BCS-IPE and BCS-MPE algorithms*

Note that as the value of hyper-parameter $a_0$ increases to infinity, the t-distribution $\text{St}\left(\mathbf{y}|\mathbf{0},\frac{b_0}{a_0}\mathbf{B},2a_0\right)$ for the evidence function $p(\mathbf{y}|\boldsymbol{\alpha},a_0,b_0)$ in (31) for BCS-IPE tends to the Gaussian PDF $\mathcal{N}\left(\mathbf{y}|\mathbf{0},\frac{b_0}{a_0}\mathbf{B}\right)$, which has the form of the evidence $p(\mathbf{y}|\boldsymbol{\alpha},\beta) = \mathcal{N}(\mathbf{y}|0,\beta^{-1}\mathbf{B})$ in Eq. (11) for BCS-MPE, which has the optimal value $\hat{\beta}^{-1} = \frac{\mathbf{y}^T\mathbf{B}^{-1}\mathbf{y}}{K-2}$ from (20). For BCS-IPE, the optimal value of the ratio $\frac{b_0}{a_0}$ is given by (32):

$$\frac{\tilde{b}_0}{\tilde{a}_0} = \frac{1}{K}\mathbf{y}^T\mathbf{B}^{-1}\mathbf{y} \tag{37}$$

where $\tilde{b}_0/\tilde{a}_0$ is close to the optimal value of $\hat{\beta}^{-1}$ for BCS-MPE for larger $K$ values. Thus, as $a_0 \to \infty$, the evidence functions for the BCS-IPE and BCS-MPE algorithms are essentially the same for large $K$ values and, therefore, so are their optimal values of the hyper-parameters $\boldsymbol{\alpha}$. However, as we have already argued, for BCS-IPE, smaller values of $a_0$ should give better performance, and the numerical results given later support this conclusion because BCS-IPE outperforms BCS-MPE.

Finally, we compare the posterior PDF over $\mathbf{w}$ in (12) and (28), which are used in BCS-MPE and BCS-IPE, respectively. It is seen that when $a_0 \to \infty$, the Student's t-distribution in (28) approaches the Gaussian PDF in (12):

$$p(\mathbf{w}|\mathbf{y},\boldsymbol{\alpha},a_0,b_0) = \mathcal{N}\left(\mathbf{w}|\mathbf{C}^{-1}\boldsymbol{\Theta}^T\mathbf{y},\frac{b'_0}{a'_0-1}\mathbf{C}^{-1}\right) = \mathcal{N}(\mathbf{w}|\boldsymbol{\mu},\boldsymbol{\Sigma}) = p\left(\mathbf{w}|\mathbf{y},\boldsymbol{\alpha},\hat{\beta}(a_0,b_0)\right) \tag{38}$$

because the covariance matrix in (29b) is:

$$\mathbf{Cov}(\mathbf{w}|\mathbf{y},\boldsymbol{\alpha},a_0,b_0) = \frac{b'_0}{a'_0-1}\mathbf{C}^{-1} = \hat{\beta}(a_0,b_0)^{-1}\mathbf{C}^{-1} = \boldsymbol{\Sigma} \tag{39}$$

If we use the optimal value $\tilde{b}_0 = \frac{\tilde{a}_0}{K}\mathbf{y}^T\mathbf{B}^{-1}\mathbf{y}$ from (32) and let $K$ be large, then from (17) with $\tilde{a}_0 \to \infty$:

$$\hat{\beta}(\tilde{a}_0,\tilde{b}_0)^{-1} \to \frac{1}{K}\mathbf{y}^T\mathbf{B}^{-1}\mathbf{y} \approx \frac{(1-2/K)^{-1}}{K}\mathbf{y}^T\mathbf{B}^{-1}\mathbf{y} = \hat{\beta}(\hat{a}_0,\hat{b}_0) \tag{40}$$

where $\hat{\beta}(\hat{a}_0,\hat{b}_0)$ is $\hat{\beta}$ in (20). Thus, the posterior means for $\mathbf{w}$ in BCS-IPE and BCS-MPE are the same and their posterior covariance matrices are $\hat{\beta}(\tilde{a}_0, \tilde{b}_0)^{-1} \mathbf{C}^{-1}$ and $\hat{\beta}(\hat{a}_0, \hat{b}_0)^{-1} \mathbf{C}^{-1}$, respectively, which are close for large $K$ values.

We conclude that as $a_0 \to \infty$, the results from the BCS-IPE algorithm are essentially the same as those from the BCS-MPE algorithms. However, we have found that using heavier-tailed multivariate Student-t distribution obtained by integration over $\beta$ and with smaller $a_0$ (larger uncertainty in the PDF over $\beta$) reconstructs signals with greater robustness and quantifies posterior uncertainty more effectively, especially when the original signal $\mathbf{x}$ is approximately sparse, which can be viewed as a sparse signal with noise.

We now examine the relative sparseness of the signal models from BCS-MPE and BCS-IPE. By comparing $\tilde{\gamma}_n$ and $\hat{\gamma}_n$ in (35) and (22), respectively, it can be deduced that $\tilde{\gamma}_n < \hat{\gamma}_n$ for any finite $a_0$, with equality as $a_0 \to \infty$. Let us examine the difference $\hat{\alpha}_n^{-1} - \tilde{\alpha}_n^{-1}$ using (34) and (21):

$$\hat{\alpha}_n^{-1} - \tilde{\alpha}_n^{-1} = \begin{cases} 0, & \text{if } \hat{\gamma}_n \leq 0 \\ \hat{\gamma}_n^{-1} & \text{if } \hat{\gamma}_n > 0 \text{ and } \tilde{\gamma}_n \leq 0 \\ \hat{\gamma}_n^{-1} - \tilde{\gamma}_n^{-1} & \text{if } \tilde{\gamma}_n > 0 \end{cases} \tag{41}$$

Therefore, the MAP estimates $\tilde{\alpha}_n$ from BCS-IPE are always smaller than MAP estimates $\hat{\alpha}_n$ from BCS-MPE when the $n^{\text{th}}$ basis vector is in included in both signal models. This observation implies that the posterior mean $\mathbf{E}(\mathbf{w}|\mathbf{y}, \tilde{\boldsymbol{\alpha}})$ obtained by the BCS-IPE method will tend to be more sparse than $\mathbf{E}(\mathbf{w}|\mathbf{y}, \hat{\boldsymbol{\alpha}}, \hat{\beta})$ from BCS-MPE when $a_0$ is finite.

For finite $a_0$, the posterior variances from $\mathbf{Cov}(\mathbf{w}|\mathbf{y}, \tilde{\boldsymbol{\alpha}})$ will be larger than those from $\mathbf{Cov}(\mathbf{w}|\mathbf{y}, \hat{\boldsymbol{\alpha}}, \hat{\beta})$ because $\tilde{\alpha}_n$ is smaller than $\hat{\alpha}_n$. Also, the posterior uncertainty for the prediction-error precision parameter $\beta$ is built into the posterior PDF $p(\mathbf{w}|\mathbf{y}, \tilde{\boldsymbol{\alpha}})$ for BCS-IPE where it has been incorporated by integrating over all possible value of $\beta$. For any suboptimal reconstructed signals, where there are too many $\alpha_n's$ non-zero compared with the optimal reconstruction, the posterior uncertainty $p(\mathbf{w}|\mathbf{y}, \boldsymbol{\alpha})$ for the suboptimal signal models will be larger, because the posterior PDF for $\beta$ is diffuse with large uncertainty. However, for the optimal signal model which corresponds to a single pronounced global maximum of the evidence function, the posterior PDF for $\beta$ will concentrated around its global maximum, so the uncertainty from $\text{cov}(\mathbf{w}|\mathbf{y}, \tilde{\boldsymbol{\alpha}})$ will tend to be smaller. These arguments give further support for the choice of $a_0 = 1$ in the BCS-IPE algorithm to give better performance in the signal reconstruction, where larger posterior variances correspond to sub-optimal signal models coming from incorrect convergence to a local maximum of the evidence.

## 3. EXPERIMENTS WITH REAL SHM SIGNALS

In this section we present results from applying our proposed BCS methods to real SHM signals and we compare our results with those from some of the state-of-the-art algorithms for CS signal reconstruction.

### *3.1. SHM accelerometer data from the Tianjin Yonghe Bridge*

The first application is to accelerometer data from the SHM system on Tianjin Yonghe Bridge (Li et al., 2013) (Figure 1), which is one of the earliest cable-stayed bridges constructed in the mainland of China. An acceleration time history of length 512 seconds and sample frequency 100 Hz was selected which came from a accelerometer installed on the deck of the main span (see Figure 2). The discrete-time signal consists of 51200 samples with a sample frequency of 100 Hz. Using the Haar wavelet transform, the wavelet coefficients of the acceleration signal are computed and shown in Figure 3(a), which reveals that the strict sparseness level of the wavelet coefficient vector is low: only 0.5% of the components have magnitude exactly zero. However, the effective sparseness level is quite high: the 16% and 7% largest magnitude components contain 80% and 60%, respectively, of the energy of the signal (in terms of the sum of square values).

We investigate the performance of the proposed BCS reconstruction algorithms by dividing the signal in Figure 2 into 100 segments of length $N = 512$ and compressing the signal in each segment by projection using the same sample of a zero-mean Gaussian random projection matrix $\mathbf{\Phi} \in \mathbb{R}^{K \times N}$ to get the compressed data $\mathbf{y}$, as in (2). For the matrix $\mathbf{\Theta} = \mathbf{\Phi}\mathbf{\Psi}$ in (3), $\mathbf{\Psi}$ is the discrete orthonormal Haar wavelet basis matrix constructed using the MATLAB routine at http://gtwavelet.bme.gatech.edu/. For a real CS accelerometer, we would obtain data already in a compressed form, where the projection arithmetic is integrated with the analog-to-digital converter in the sensor itself, so the actual signal, denoted $\bar{\mathbf{x}}$ and its wavelet coefficients $\bar{\mathbf{w}} = \mathbf{\Psi}^T \bar{\mathbf{x}}$, would be unknown. We label the reconstruction problem for the compressed measurement $\mathbf{y} = \mathbf{\Phi}\bar{\mathbf{x}}$ as Case 1. This is an example of a non-sparse case that is of interest when compressing and decompressing SHM signals.

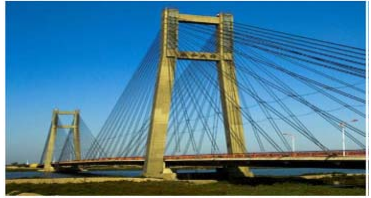

**Figure 1** Photo of the Tianjin Yonghe Bridge

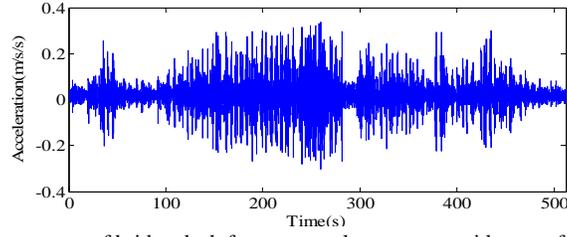

**Figure 2** Acceleration response of bridge deck from an accelerometer at mid-span of Tianjin Yonghe Bridge.

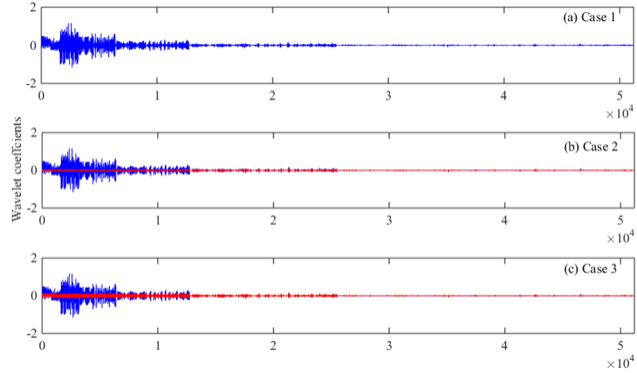

**Figure 3** Wavelet coefficients of the acceleration data in Figure 2 for Cases 1-3 using Haar wavelet basis (The red spikes in (b) and (c) denote the small magnitude components that are discarded for the two de-noised cases and the blue spikes are the retained components).

We also consider another two cases, labelled Case 2 and Case 3, in order to investigate the performance of the proposed BCS reconstruction algorithms for more highly sparse signals. We use hard threshold de-noising to set to zero the smallest coefficients with magnitudes containing 20% and 40% of the energy from the wavelet coefficients shown in Figure 3(a), giving 84% and 93%, respectively, of the wavelet coefficients exactly zero, for Case 2 and Case 3, respectively. The de-noised wavelet coefficients vector for Case 2 and Case 3 are shown in Figures 3(b) and 3(c), respectively. We then use the inverse wavelet transform and divide the obtained time series into 100 segments of length $N = 512$. For each segment, the true signal $\bar{\mathbf{x}}_d = \mathbf{\Psi}\bar{\mathbf{w}}_d$ (Cases 2 and 3) is compressed using the same projection matrix $\mathbf{\Phi}$ that was used in Case 1 to get the compressed measurement vector $\mathbf{y}_d = \mathbf{\Phi}\bar{\mathbf{x}}_d$.

For data decompression, we run the Bayesian CS algorithms to produce a probabilistic description of the reconstructed coefficients $\mathbf{w}$ and $\mathbf{w}_d$ with mean $E(\mathbf{w}|\mathbf{y})$ and $E(\mathbf{w}_d|\mathbf{y}_d)$ from compressed measurements $\mathbf{y}$ and $\mathbf{y}_d$, respectively. The corresponding uncertain reconstructed acceleration signals $\mathbf{x}$ and $\mathbf{x}_d$ are then obtained by wavelet transforms: $\mathbf{x} = \mathbf{\Psi}\mathbf{w}$ and $\mathbf{x}_d = \mathbf{\Psi}\mathbf{w}_d$. In the results shown later, the *strict* *reconstruction-error measures* for the signals recovered from the compressed measurements $\mathbf{y}$ (Case 1) and $\mathbf{y}_d$ (Cases 2 and 3) are defined by:

$$\mathbf{RE} = \|\bar{\mathbf{w}} - \mathbf{E}(\mathbf{w}|\mathbf{y})\|_2^2 / \|\bar{\mathbf{w}}\|_2^2 = \|\mathbf{\Psi}\bar{\mathbf{w}} - \mathbf{\Psi} \cdot \mathbf{E}(\mathbf{w}|\mathbf{y})\|_2^2 / \|\mathbf{\Psi}\bar{\mathbf{w}}\|_2^2 = \|\bar{\mathbf{x}} - \mathbf{E}(\mathbf{x}|\mathbf{y})\|_2^2 / \|\bar{\mathbf{x}}\|_2^2 \qquad (42)$$

$$\mathbf{RE}_d = \|\bar{\mathbf{w}}_d - \mathbf{E}(\mathbf{w}_d|\mathbf{y}_d)\|_2^2 / \|\bar{\mathbf{w}}_d\|_2^2 = \|\mathbf{\Psi}\bar{\mathbf{w}}_d - \mathbf{\Psi} \cdot \mathbf{E}(\mathbf{w}_d|\mathbf{y}_d)\|_2^2 / \|\mathbf{\Psi}\bar{\mathbf{w}}_d\|_2^2 = \|\bar{\mathbf{x}}_d - \mathbf{E}(\mathbf{x}_d|\mathbf{y}_d)\|_2^2 / \|\bar{\mathbf{x}}_d\|_2^2 \quad (43)$$

where we have used the orthonormality of the wavelet basis, so $\mathbf{\Psi}^T\mathbf{\Psi} = \mathbf{I}_N$. It is known from wavelet threshold denoising (Donoho and Johnstone, 1994) that the wavelet transform, especially the orthogonal wavelet transform, has strong denoising coherency, i.e., it tends to concentrate the energy of useful information on the larger wavelet coefficients and distribute the energy of the "noise" over the whole vector of wavelet coefficients. Therefore, we expect the wavelet coefficients with larger amplitudes to constitute the desirable part of the original signal and be sparsely distributed. The CS reconstruction of the effective wavelet vector $\bar{\mathbf{w}}_s$ may therefore be of more interest; it contains only the $T$ wavelet coefficients with magnitudes significantly larger than the "noise" background. We introduce the index vector $\mathbf{id}$ to indicate the locations of the $T$ non-zero components in $\bar{\mathbf{w}}_s$. We also calculate the *effective reconstruction-error measures* for reconstructed $\mathbf{w}_s$ and corresponding time domain signal $\mathbf{x}_s = \mathbf{\Psi}\mathbf{w}_s$ as:

$$\mathbf{RE}_s = \|\bar{\mathbf{w}}_s - \mathbf{E}(\mathbf{w}_s|\mathbf{y})\|_2^2 / \|\bar{\mathbf{w}}_s\|_2^2 = \|\bar{\mathbf{x}}_s - \mathbf{E}(\mathbf{x}_s|\mathbf{y})\|_2^2 / \|\bar{\mathbf{x}}_s\|_2^2 \qquad (44)$$

where the nonzero components of $\mathbf{w}_s$ consist of the corresponding components in $\mathbf{w}$ for fixed $\mathbf{id}$ (it is known in the test). For a proper comparison later, the effective reconstruction-error measures $(\mathbf{RE}_d)_s$ for Cases 2 and 3 are also computed using the same strategy.

In Figure 4, an example of reconstructed wavelet coefficients using Algorithm BCS-IPE on the first time segment of the Tianjin Yonghe Bridge signal is shown. The strict reconstruction error $\mathbf{RE}$ for Cases 1, 2 and 3 are 0.3163, $8.5 \times 10^{-5}$ and $1.2 \times 10^{-5}$, respectively. The reconstruction error for Case 1 is too large to be acceptable for real applications. However, we observe that most of the wavelet coefficients in $\bar{\mathbf{w}}$ are minor. If we choose the first $1/16$ ($T = 32$) and $1/4$ ($T = 128$) wavelet coefficients in $\bar{\mathbf{w}}$ when listed in decreasing magnitude to constitute the non-zero coefficients in $\bar{\mathbf{w}}_s$, then the effective reconstruction-error measures $\mathbf{RE}_s$ quantified by (44) become much smaller with values of 0.1142 and 0.1923, respectively, for Case 1, demonstrating that much more accurate reconstruction can be achieved for wavelets coefficients with larger amplitudes.

An important advantage of the BCS-IPE method is that it establishes the posterior distribution of the unknown wavelet coefficients without any user parameter setting. This distribution can be used to provide effective uncertainty quantification for the reconstructed wavelet coefficients as shown in Figure 4 (d)-(f), where the error-bars are defined as ± one posterior standard deviation, computed from the diagonal elements of the covariance matrix $\boldsymbol{\Sigma}$ of $\mathbf{w}$. It is seen that the error bars in Figure 4(d) are obviously larger compared with those in Figures 4(e) and 4(f), which indicates that the reconstruction confidence for approximately-sparse signals is smaller than for the sparse signals. This is consistent with the fact that it is unlikely to give an exact reconstruction in this case.

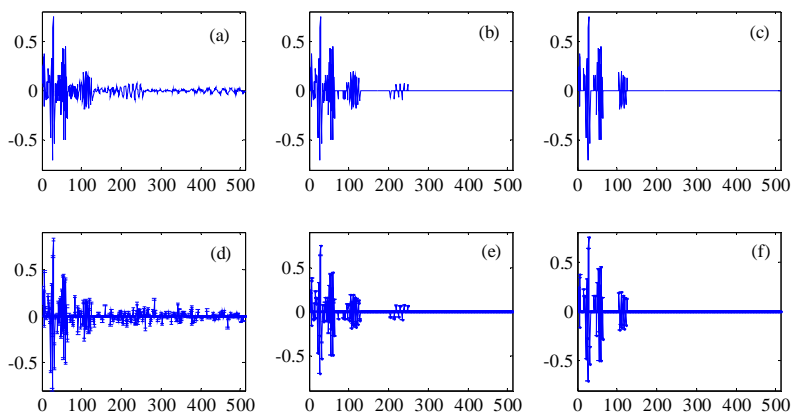

**Figure 4.** The wavelet coefficients of (a) original (Case 1), (b) 20% de-noised (Case 2), and (c) 40% de-noised (Case 3) signals, from the first time segment of length $N = 512$ of the acceleration signal from the SHM system on the Tianjin Yonghe Bridge; (d), (e) and (f) show reconstruction results for Cases 1, 2 and 3, respectively, using algorithm BCS-IPE with the number of compressed measurements $K = 300$. The closely-spaced upper and lower short horizontal lines for each reconstructed wavelet coefficient delineate the error bars, with the centers of the vertical lines corresponding to the posterior mean.

In Figures 5-7, the proposed Bayesian algorithms (BCS-MPE and BCS-IPE) are compared with respect to two published Bayesian CS algorithms, BCS (Ji et al., 2008) and BCS-Laplace (Babacan et al., 2010), using the matlab codes for them that were downloaded from http://people.ee.duke.edu/~lcarin/BCS.html and http://ivpl.eecs.northwestern.edu/research/topics/compressive-sensing, respectively, and three state-of-the-art deterministic CS reconstruction algorithms: BP (Candes et al., 2006) using the l1-magic package at http://www-stat.stanford.edu/~candes/l1magic/, GPSR (Figueiredo et al., 2007) with the matlab code downloaded from http://www.lx.it.pt/~mtf/GPSR/, and AIHT (Blumensath, 2012) with the code from http://users.fmrib.ox.ac.uk/~tblumens/sparsify/sparsify.html. For all these algorithms, the required parameters are set according to their algorithm default setups.

For the purpose of examining the signal reconstruction performance for various compression ratios ($CR$), we vary the number of compressed measurements $K$ from 170 to 470 (compression ratios $CR$ of $N/K$ from 1.09 to 3.01, where $N = 512$) for Cases 1 (Figure 5) and 2 (Figure 6) and from 80 to 470 (compression ratios $CR$ from 1.09 to 6.4) for Case 3 (Figure 7). Different thresholds (0.01, 0.02, 0.05, 0.10, and 0.20) of acceptable reconstruction errors $\mathbf{RE}, \mathbf{RE}_d, \mathbf{RE}_s$ and $(\mathbf{RE}_d)_s$, as quantified in (42-44), are also employed to present the rates of acceptable performance for the 100 time segments, based on the results using different numbers of compressed measurements $K$. In these experiments, the first $1/16$ and $1/4$ largest magnitude wavelet coefficients in $\bar{\mathbf{w}}$ (for Case 1) and $\bar{\mathbf{w}}_d$ (for Cases 2 and 3) are selected to constitute non-zero coefficients in $\bar{\mathbf{w}}_s$ and $(\bar{\mathbf{w}}_d)_s$, respectively. As expected, increases in $CR$ beyond a critical value correspond to a decrease in the rates of acceptable performance for all of these methods, as observed in Figures 5-7, and the increase in sparseness level from Case 1 to Case 2 and Case 2 to Case 3 corresponds to an increase in this critical value of CR.

Comparing acceptance rates in different columns of Figure 5 for different levels of reconstruction error in Case 1, it is seen that more reconstructions are acceptable if judged by only the largest effective wavelet coefficients, which is consistent with the conclusion found from Figure 4. Figure 5(a) shows that when the compression ratio $CR$ is smaller than 1.5, all effective reconstruction error measures for the $1/16$ largest magnitude wavelet coefficients are smaller than 0.1 for BP and all of the Bayesian CS methods.

From the comparison of the different algorithms in Figures 5-7, it is seen that the proposed BCS-IPE algorithm outperforms all other methods in terms of acceptable performance. When the threshold is set to 0.01 for Case 1 (Figure 5(a), (b) and (c)), there are no acceptable reconstructions produced by any of the algorithms. On the other hand, for Case 2 (Figure 6) and Case 3 (Figure 7), all reconstructions are nearly perfect (RE<0.01) for BCS-IPE when the compression ratios are smaller than 2.3 and 3.0, respectively. In addition, the proposed BCS-IPE method has the advantage that all parameters are solely learned from the data adaptively and automatically, thereby avoiding user intervention to set parameters related to signal sparseness, noise levels, etc, which is needed in the deterministic algorithms. It is concluded that the proposed BCS-IPE algorithm provides the best overall performance among all the presented methods for approximately sparse signals.

It is interesting to note that the improvements in the reconstruction accuracy of BCS-IPE compared with BCS-MPE, are minor for signals in Case 3 (Figure 7). This is expected because that the integration over the prediction error precision $\beta$ to account for its posterior uncertainty does not gain much for a very sparse signal, which has a signal model class $\mathcal{M}(\boldsymbol{\alpha}, \beta)$ that has a distinct global peak of the evidence function $p(\mathbf{y}|\boldsymbol{\alpha}, \beta)$, so the Laplace approximation using only the MAP value of $\beta$ is accurate.

Compared with the deterministic CS reconstruction algorithms (BP, GPSR and AIHT), the Bayesian CS methods have the advantage that they quantify the posterior uncertainty or confidence for signal reconstructions. By implementation of the four BCS algorithms for Cases 1-3 over the first 50 time segments, the reconstruction-error measures and the averages of the posterior standard deviations over all nonzero reconstructed wavelet coefficients are shown in Figure 8. In the test, we set the number of measurement $K = 230, 200$ and 100, for Cases 1, 2 and 3, respectively.

Perfect correspondence is observed between the occurrence of the larger averages of the posterior standard deviations and the occurrence of the suboptimal reconstructions in Cases 2 (Figure 8(c-d)(iv)) and 3 (Figure 8(e-f)(iv)) for BCS-IPE, although there is no such correspondence observed for Case 1 because the sparseness levels are too low in the studied signals and no perfect reconstruction is achieved. For the other three BCS algorithms, the correspondence between suboptimal reconstructions and larger averages of posterior standard deviations cannot be found and the quantification of the signal reconstruction uncertainty is confusing, that is, there is no correlation between poor signal reconstructions and large posterior uncertainty.

Together, these observations demonstrate that the proposed BCS-IPE provides the best overall performance among all methods considering reconstruction robustness and posterior uncertainty quantification. Therefore, implementation of the CS technique in the sensors, along with the reconstruction algorithm BCS-IPE in the central processing unit (CPU), is suggested for wireless structural health monitoring. However, a practical issue that needs to be addressed is that data loss may occur during wireless transmission from a sensor to the CPU (Meyer et al., 2010). In Figure 9, the signal reconstruction performance is investigated when there are $K_m$ data points lost in $\mathbf{y} = \boldsymbol{\Phi}\bar{\mathbf{x}}$ and the received data vector $\mathbf{y}_l$ contains only $K_l = K - K_m$ data points. Signal recovery is essentially the same as the data decompression in CS: the corresponding $K_m$ rows of the projection matrix $\boldsymbol{\Phi} \in \mathbb{R}^{K \times N}$ are discarded to get a new matrix $\boldsymbol{\Phi}_l \in \mathbb{R}^{K_l \times N}$ with a smaller number of rows, that is, the received compressed measurement vector $\mathbf{y}_l \in \mathbb{R}^{K_l \times 1}$ is effectively produced by linear projections of the original signal $\bar{\mathbf{x}}$ using matrix $\boldsymbol{\Phi}_l$, $\mathbf{y}_l = \boldsymbol{\Phi}_l \bar{\mathbf{x}}$.

In a wireless sensor network, the data packets, each of which contains a certain number of data points, are transmitted one by one, and all the data points in a lost packet will be missing. In our data-recovery experiments, we assume four sampling points are included in each data packet and therefore 128 data packets are required for an uncompressed measurement **y** ($K = N = 512$). For the purpose of examining the signal reconstruction performance of BCS-IPE for different data loss rates, we vary the number of lost data packets from 1 to 26 (data loss rate of $K_l/K$ from 0.78% to 20.31%). The lost data packets are selected randomly among the 128 candidates, and we execute the same experiment 100 times and report the overall reconstruction performance. As in Figures 5-7, Figure 9 shows different thresholds (0.02, 0.05, 0.10, and 0.20) of acceptable reconstruction errors to denote the rates of acceptable performance for the 100 runs for each possible data loss rate. It is observed that almost all reconstructions are acceptable if the threshold of acceptable strict reconstruction error measures is set to be 0.1 (Figure 9 (c)), as long as the data loss rates are smaller than 9%. For relatively larger magnitude wavelet coefficients, it is seen that all reconstructions have effective reconstruction errors smaller than 0.05 and 0.10 when investigating the $1/16$ and $1/4$ largest magnitude wavelet coefficients, respectively, even for 20% data loss rate (Figure 9 (a) and (b)). BCS-IPE is therefore a promising algorithm for automated recovery of any data lost during wireless transmission, which can be used to guard against data loss even if the signal is not sparse in any basis.

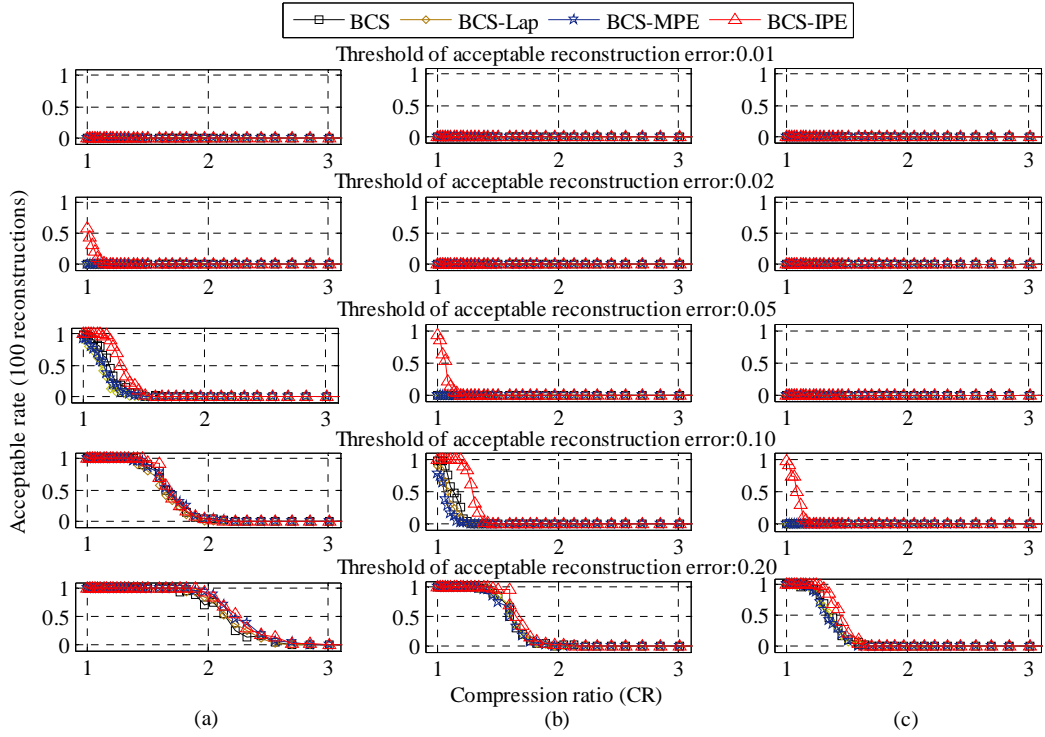

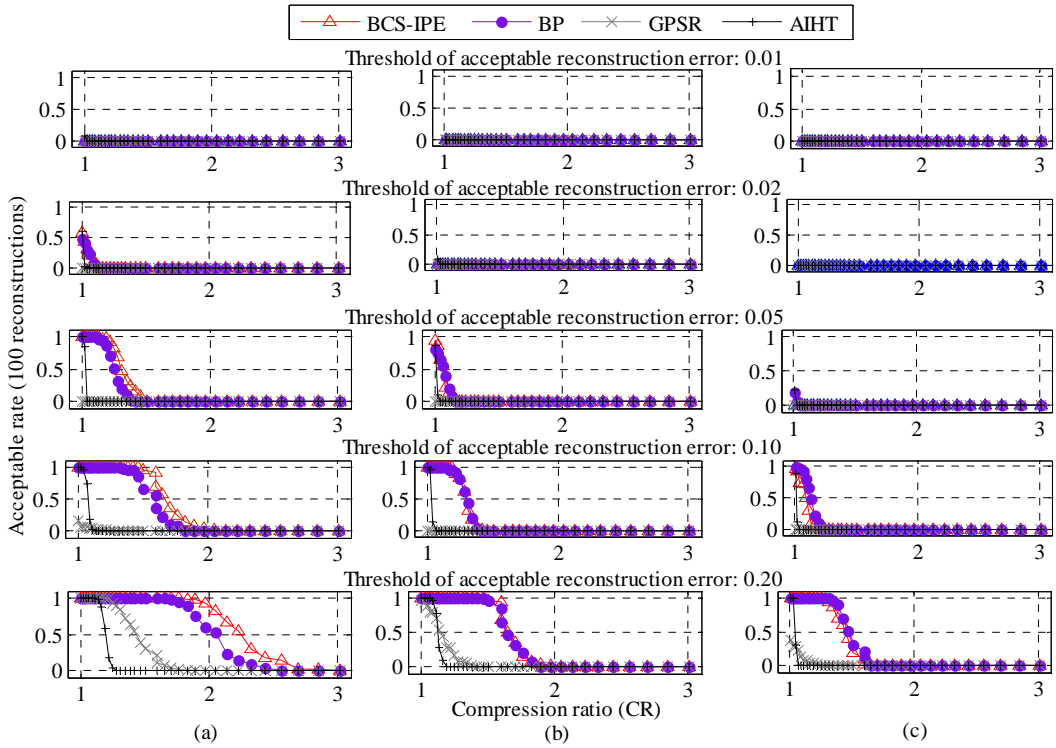

**Figure 5:** Case 1 (Original signal from Tianjin Yonghe Bridge): Relation of compression ratio and the rate of acceptable reconstruction error for 5 different thresholds for the reconstruction errors of (a) 1/16 largest coefficients; (b) 1/4 largest coefficients; and (c) all coefficients of the original wavelet coefficient vector.

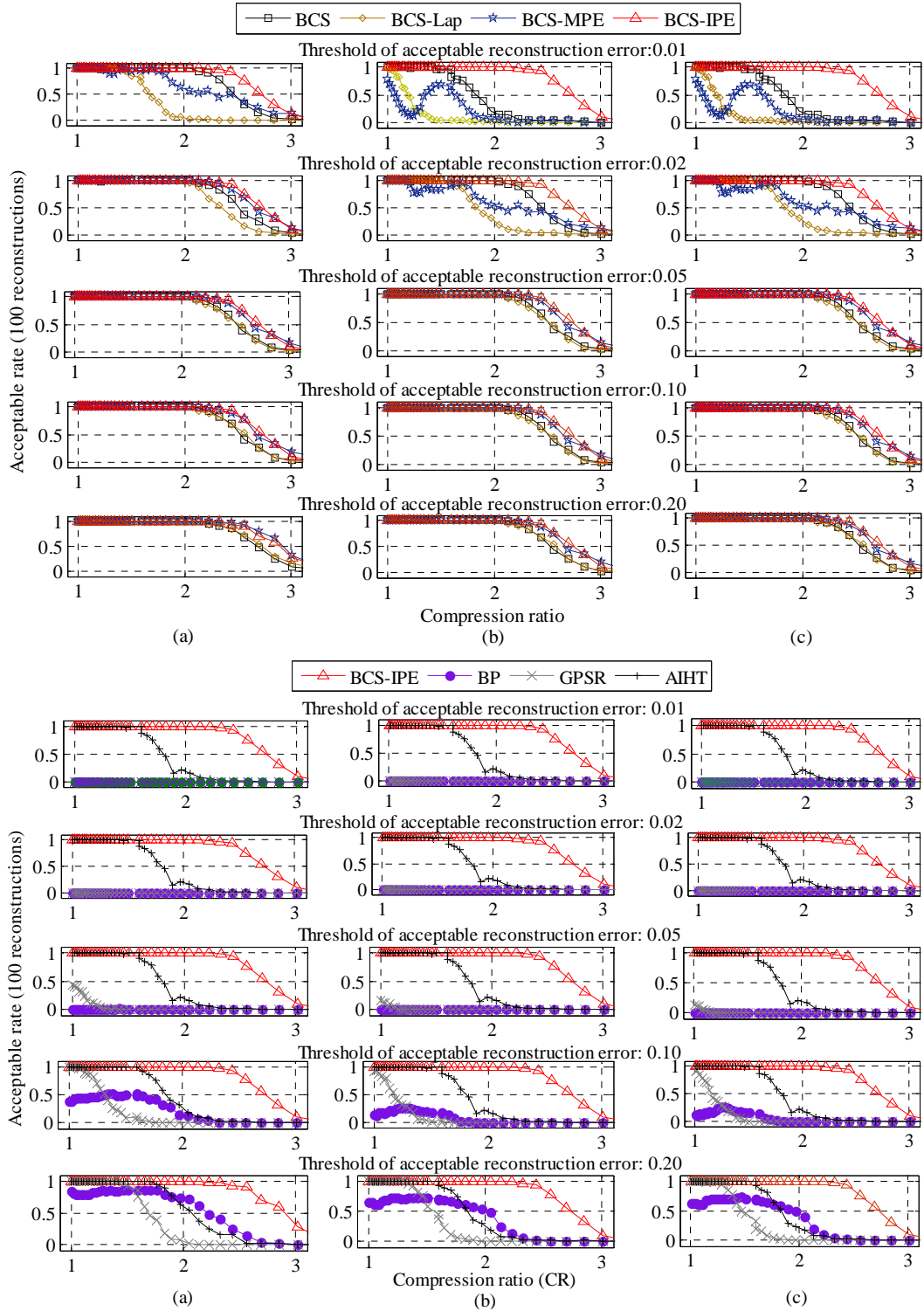

**Figure 6:** Case 2 (20% de-noised signal from Tianjin Yonghe Bridge): Relation of compression ratio and the rate of acceptable reconstruction error for 5 different thresholds for the reconstruction errors of (a) 1/16 largest coefficients; (b) 1/4 largest coefficients; and (c) all coefficients of the original wavelet coefficient vector.

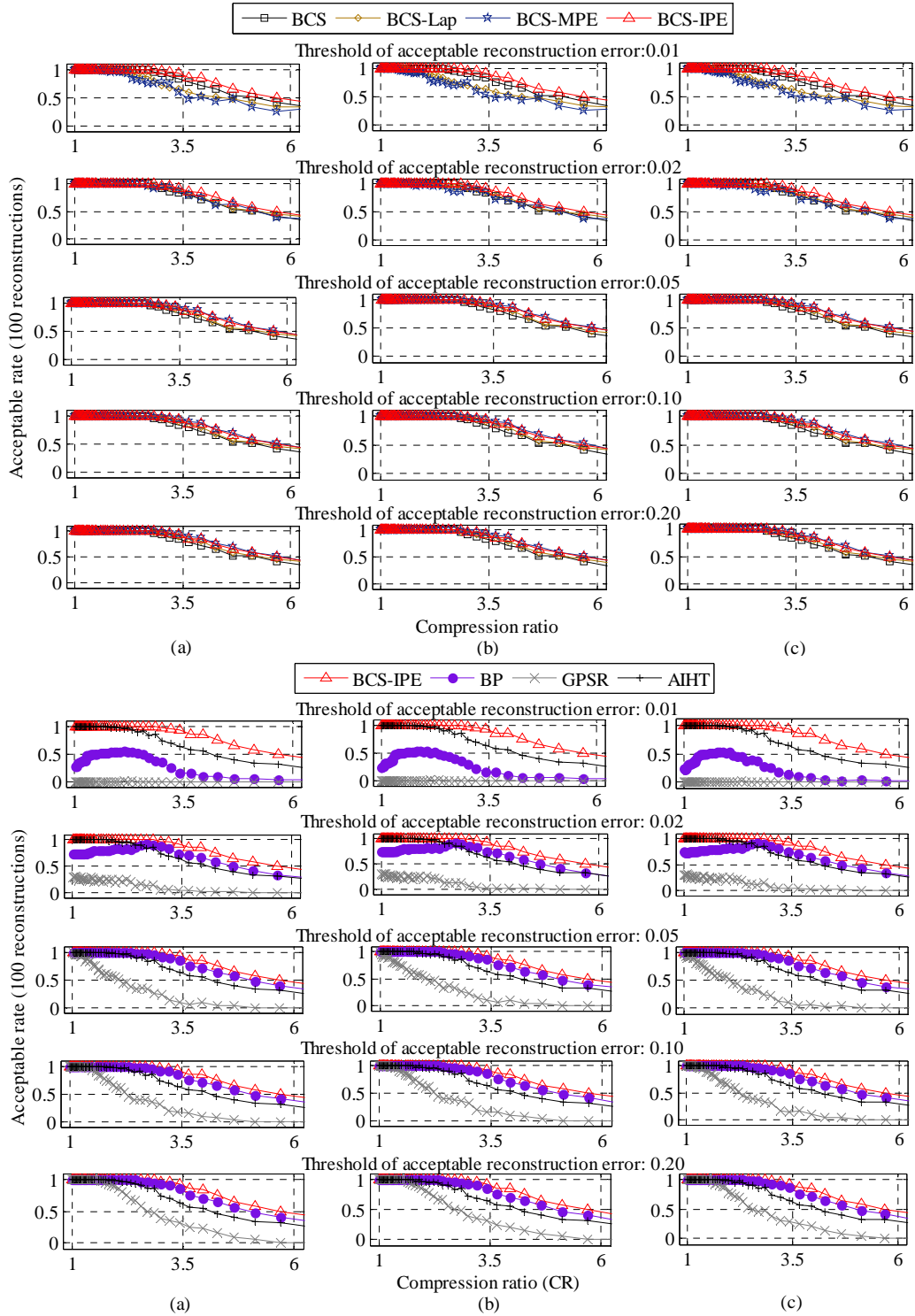

**Figure 7:** Case 3 (40% de-noised signal from Tianjin Yonghe Bridge): Relation of compression ratio and the rate of acceptable reconstruction error for 5 different thresholds for the reconstruction errors of (a) 1/16 largest coefficients; (b) 1/4 largest coefficients; and (c) all coefficients of the original wavelet coefficient vector.

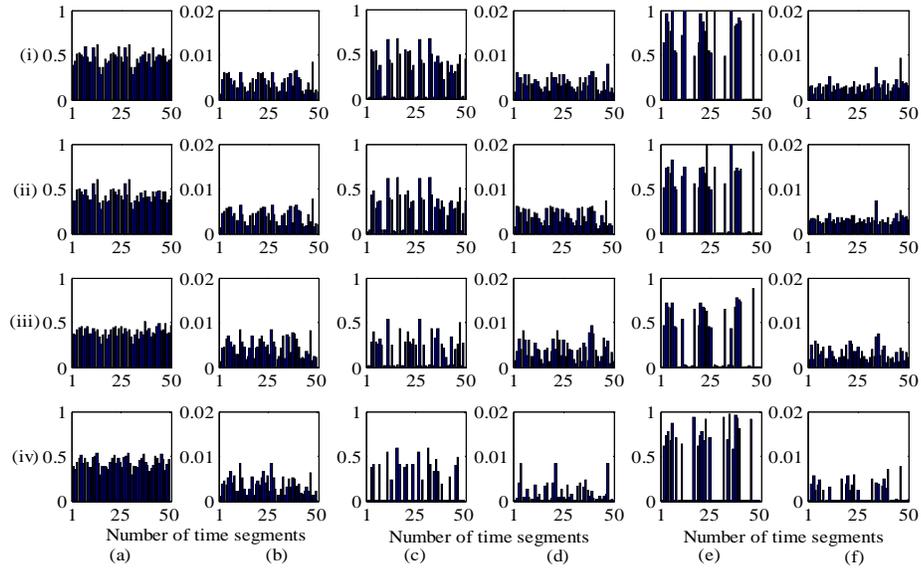

**Figure 8:** Reconstruction error measures ((a),(c),(e)) and corresponding averages ((b),(d),(f)) over all nonzero posterior standard deviations for the first 50 time segments of: (a),(b): the original (Case 1, $K = 230$); (c),(d): 20% de-noised (Case 2, $K = 200$); and (e),(f): 40% de-noised (Case 3, $K = 100$) signals from Tianjin Yonghe Bridge. The BCS algorithms used are: (i) BCS; (ii) BCS-Lap; (iii) BCS-MPE; (iv) BCS-IPE.

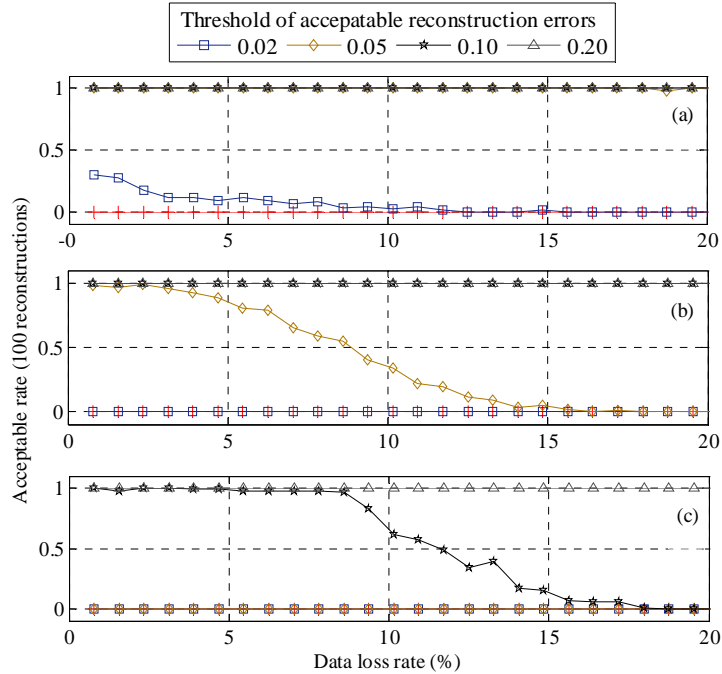

**Figure 9:** Relation between data loss rate and the rate of acceptable reconstruction error for 5 different thresholds for the reconstruction errors for BCS-IPE of (a) $1/16$ largest coefficients; (b) $1/4$ largest coefficients; and (c) all coefficients of the original wavelet coefficient vector for original signals (Case 1) from Tianjin Yonghe Bridge.

*3.2 SHM accelerometer data from the Beijing National Aquatics Center*

In this section, we investigate CS reconstruction for acceleration data from the Beijing National Aquatics Center (Figure 10), popularly called the Water Cube. It is a well-known steel space-frame structure built for the 2008 Olympics swimming facility. It holds a record for the largest ETFE (Ethylene tetrafluoroethylene) clad structure in the world. A sophisticated long-term structural health monitoring system (Ou and Li, 2010) was installed on this structure in 2008. An ambient vibration response signal of length 512 seconds (Figure 11) and sample frequency 100 Hz from one of the accelerometers is studied here.

Using the Haar wavelet transform, the wavelet coefficients of the acceleration signal are computed and shown in Figure 12(a), which reveals that the strict sparseness level of the wavelet coefficients vector is much lower than that for Tianjin Yonghe Bridge (see Figure 3) and none of the wavelet coefficients are exactly zero; this is because the acceleration data collected from the Water Cube has a more wideband frequency content. The different sparseness levels from the two structures are a consequence of the different dynamic characteristic. The effective sparseness level is not high since the 45% and 27% largest magnitude components contain 80% and 60%, respectively, of the energy of the signal (in terms of the sum of square values). We label the reconstruction problem corresponding to original signal as Case 1.

Using the same hard threshold de-noising strategy as in Section 3.1, the smallest 55% and 73% of the coefficients containing 20% and 40% of the energy, respectively, are set to zero, leaving 45% and 27% of the de-noised coefficients as the only non-zero ones. The resulting wavelet coefficients are shown in Figure 10(b) and (c). We label the reconstruction problems corresponding to the 20% and 40% de-noised signals as Case 2 and Case 3, respectively. The same setup and projection matrix Φ used in Section 3.1 is also employed here.

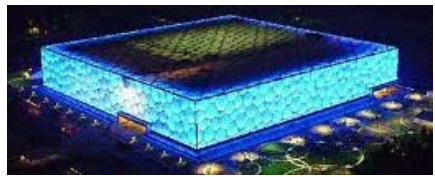

**Figure 10.** Beijing National Aquatics Center.

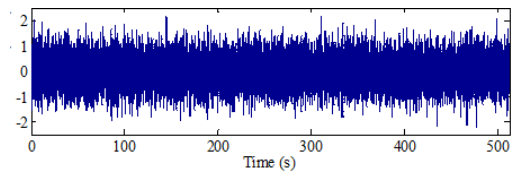

**Figure 11.** Signal from an accelerometer on the bottom chord plane of Beijing National Aquatics Center.

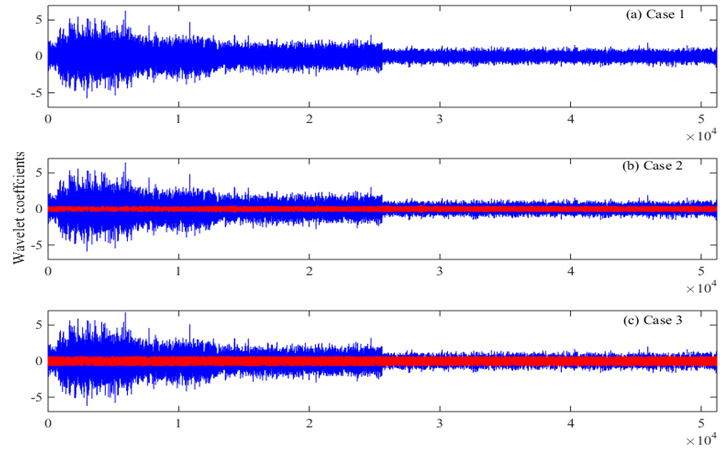

**Figure 12.** Wavelet coefficients of the acceleration signal in Figure10 for Cases 1-3 using Haar wavelet basis (The red spikes in (b) and (c) denote small magnitude components that are discarded for the two de-noised cases and the blue spikes are the retained components).

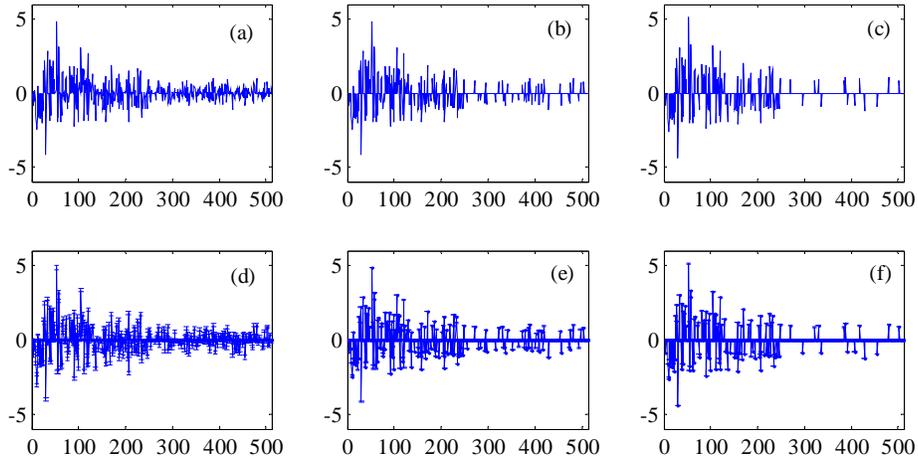

**Figure 13.** The wavelet coefficients of (a) original (Case 1), (b) 20% de-noised (Case 2), and (c) 40% de-noised (Case 3), from the first time segment of length $N = 512$ of the acceleration signal from the SHM system on the Beijing National Aquatics Center; (d) (e) and (f) show reconstruction results for Cases 1, 2 and 3, respectively, using algorithm BCS-IPE with the number of compressed measurements $K = 400$. The upper and lower short horizontal lines for each reconstructed wavelet coefficient delineate the error bars with the centers of the vertical lines corresponding to the posterior mean.

We first focus on signal reconstruction of the first time segment using the proposed algorithm BCS-IPE. In Figure 13 (a)-(c), we present the wavelet coefficients for the original (Case 1) and 20% de-noised (Cases 2) and 40% de-noised (Case 3) signals. As in Figure 4, we also present the posterior mean and error-bars of the reconstructed wavelet coefficients in Figures 13 (d)-(f). High reconstruction accuracy is again observed for BCS-IPE, especially for the larger wavelet coefficients. The corresponding posterior reconstruction uncertainty for Case 1 is much larger than those for Cases 2 and 3, indicating the confidence for the inverse model for approximately-sparse signals in Case 1 is smaller.

Similar to Figures 5, 6 and 7, Figures 14, 15 and 16 present the reconstruction performance for various methods with different acceptable thresholds for Cases 1, 2 and 3, respectively. Because of the less sparse signals here, the rates of acceptable reconstruction in Figures 14, 15 and 16 are generally much lower than those in Figures 5, 6 and 7, respectively, for the same compressive ratios. By comparing different algorithms, a similar conclusion to that in Section 3.1 is reached: BCS-IPE shows superior performance over all other algorithms, with higher acceptance rates being achieved for a given compression ratio CR.

In Figure 17, the results of the reconstruction error measure and corresponding average of the posterior standard deviations for the reconstruction of the first 50 time segments of the original and de-noised signals are given for different algorithms. For BCS-IPE, the correlation of the occurrence of smaller reconstruction errors with the occurrence of smaller averages of the posterior standard deviations of the wavelet coefficients is evident in Cases 2 (Figure 17 (c-d) (iv)) and 3 (Figure 17 (e-f) (iv)). However, there is no similar correlation for Case 1 (Figure 17 (a-b) (iv)).

In Figure 18, the performance of data loss recovery is studied for the original signal (Case 1) from the Beijing National Aquatics Center. Even though there is less sparseness in the signal compared with Tianjin Yonghe Bridge, almost all reconstructions for the 1/16 and 1/4 largest magnitude wavelet coefficients are acceptable when we set the thresholds of acceptable reconstruction errors as 0.05 and 0.10, respectively, and the data loss rate is less than 8%. These reconstruction errors are thought to be tolerable for structural modal identification and damage assessment using the reconstructed signals.

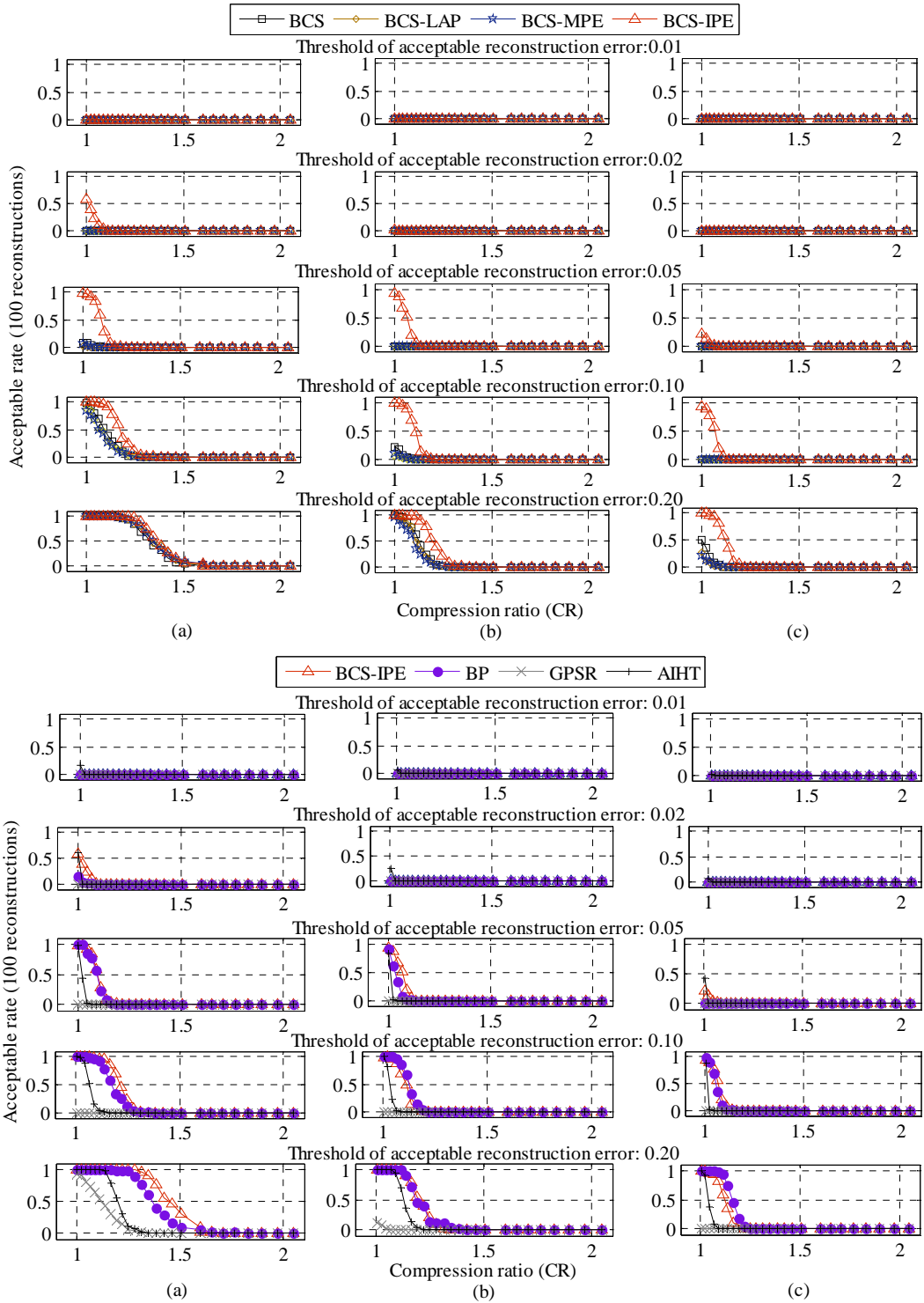

**Figure 14:** Case 1 (Original signal from Beijing National Aquatics Center): Relation of compression ratio and the rate of acceptable reconstruction error for 5 different thresholds for the reconstruction errors of (a) 1/16 largest coefficients; (b) 1/4 largest coefficients; and (c) all coefficients of the original wavelet coefficient vector.

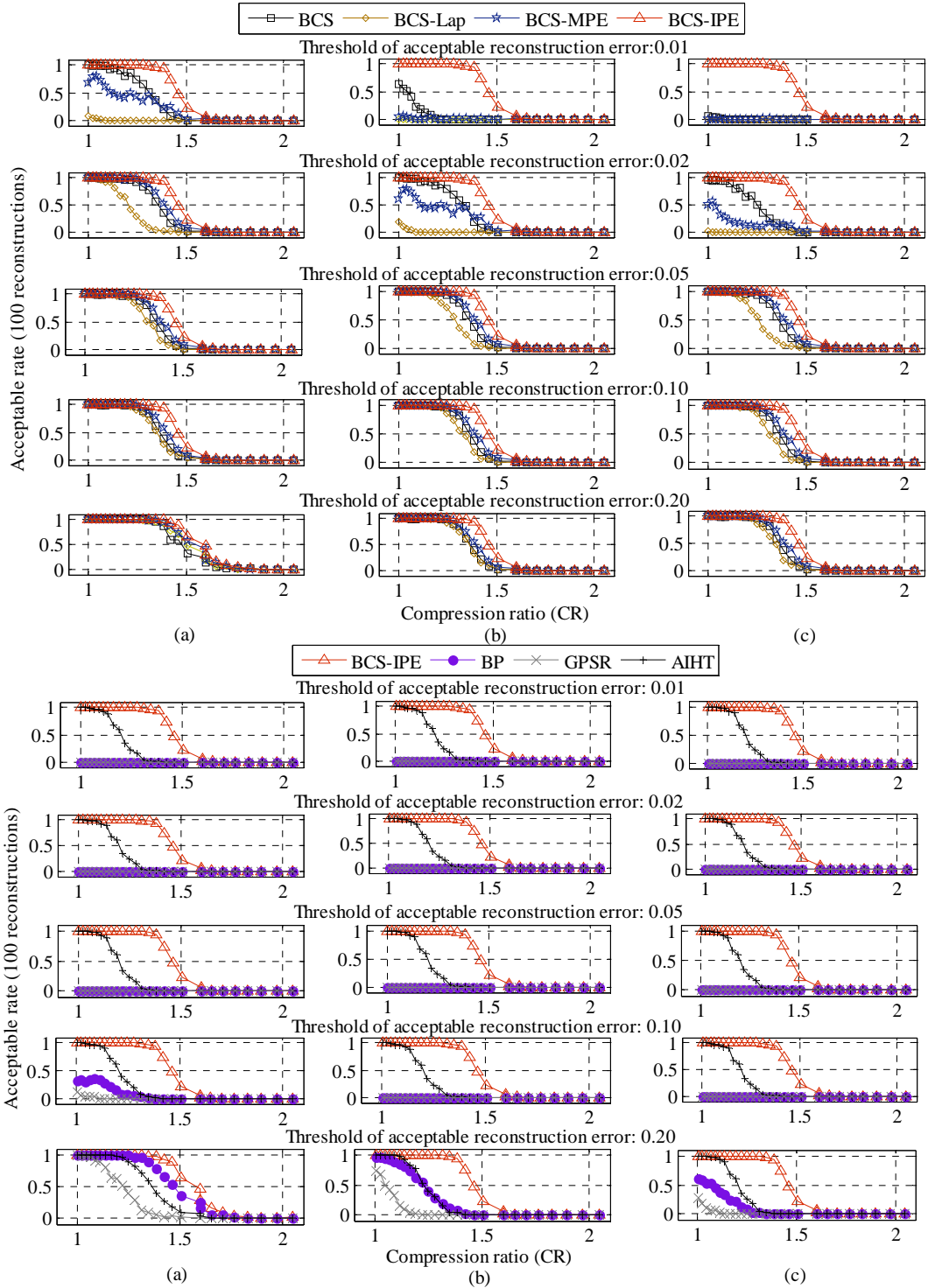

**Figure 15:** Case 2 (20% de-noised signal from Beijing National Aquatics Center): Relation of compression ratio and the rate of acceptable reconstruction error for 5 different thresholds for the reconstruction errors of (a) $1/16$ largest coefficients; (b) $1/4$ largest coefficients; and (c) all coefficients of the original wavelet coefficient vector.

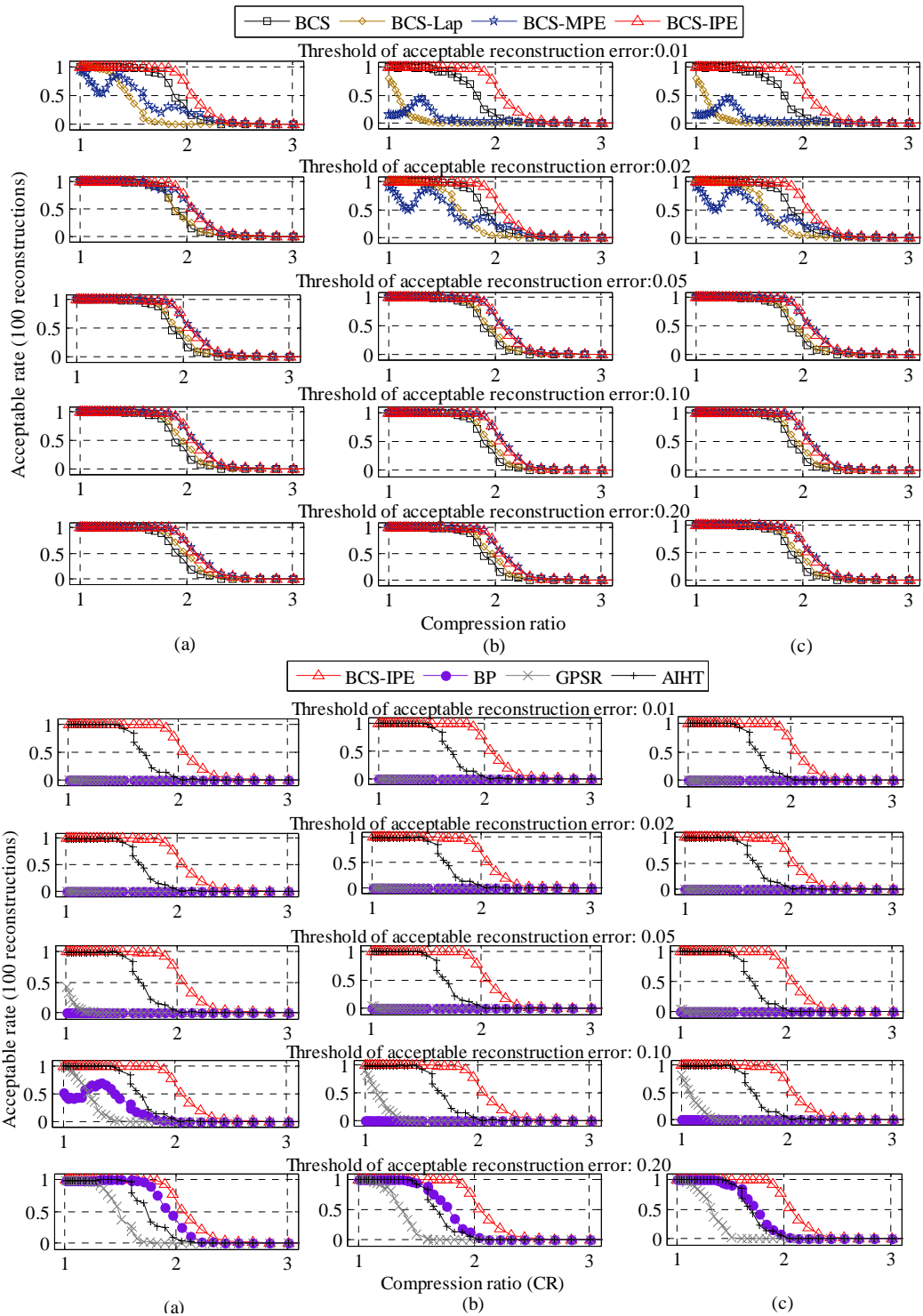

**Figure 16:** Case 3 (40% de-noised signal from Beijing National Aquatics Center): Relation of compression ratio and the rate of acceptable reconstruction error for 5 different thresholds for the reconstruction errors of (a) 1/16 largest coefficients; (b) 1/4 largest coefficients; and (c) all coefficients of the original wavelet coefficient vector.

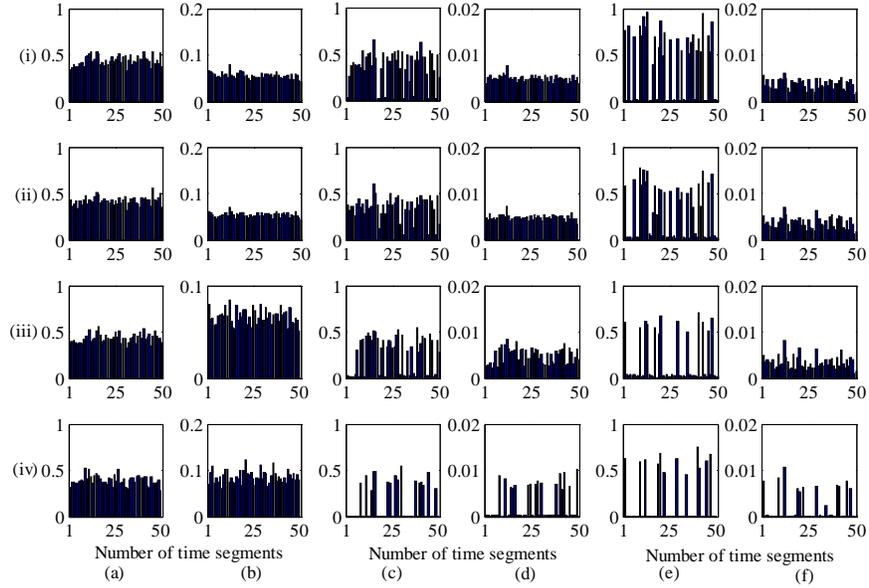

**Figure 17:** Reconstruction error measures ((a),(c),(e)) and corresponding averages ((b),(d),(f)) over all nonzero posterior standard deviations for the first 50 time segments of: (a),(b): the original (Case 1, $K = 400$); (c),(d): 20% de-noised (Case 2, $K = 360$); and (e),(f): 40% de-noised (Case 3, $K = 260$) signals from Beijing National Aquatics Center. The BCS algorithms used are: (i) BCS; (ii) BCS-Lap; (iii) BCS-MPE; (iv) BCS-IPE.

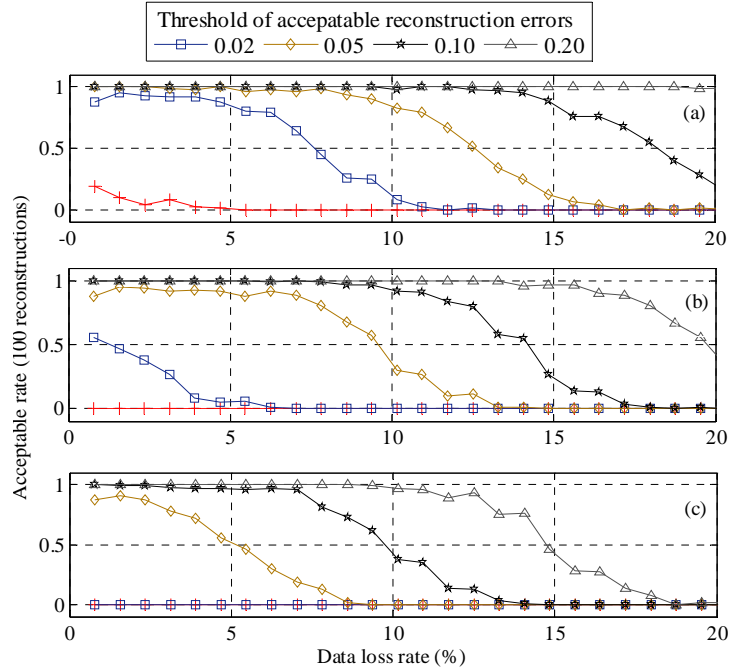

**Figure 18:** Relation between data loss rate and the rate of acceptable reconstruction error for 5 different thresholds for the reconstruction errors for BCS-IPE of (a) $1/16$ largest coefficients, (b) $1/4$ largest coefficients, and (c) all coefficients of the original wavelet coefficient vector for original signals (Case 1) from Tianjin Yonghe Bridge.

**CONCLUDING REMARKS**

Compressive sensing techniques are promising for SHM systems to increase the efficiency of wireless data transmission and for data loss recovery. Most of the existing CS techniques allow effective reconstruction of signals only when they are sufficiently sparse in terms of some orthogonal basis. However, real signals in SHM are usually only "approximately sparse", so Bayesian CS for approximately-sparse signals is explored in this research.

We present the BCS-IPE algorithm for robust treatment of the prediction error precision $\beta$ where we marginalize out this parameter to effectively account for its posterior uncertainty. The effective dimensionality (number of nonzero basis coefficients) of the signal model is determined automatically as part of the full Bayesian inference procedure, and all uncertain model parameters are estimated solely from the compressed data. The BCS-IPE algorithm produces more accurate reconstructions for approximately sparse SHM signals than the BCS-MPE algorithm based on MAP estimation of the prediction error precision $\beta$. In addition, the posterior uncertainty quantification for the signal reconstructions is more reliable for BCS-IPE, so it is more useful tool for signal reconstructions. Although the allowable compression ratios for reliable signal reconstructions are not so high for the investigated real SHM signals because of their low sparseness, they are sufficient to allow acceptable signal recovery of around 7-9% loss of data during wireless transmission.

In the set of presented experiments, the reconstruction performance of the larger coefficients in the wavelet basis domain is also investigated. It is found the reconstruction accuracy and uncertainty quantification for wavelet coefficients with large amplitudes are much better than those with small amplitudes using the proposed BCS-IPE algorithm. Furthermore, it allows acceptable signal recovery for 10-15% loss of data. Such reconstructed signals may therefore allow reliable modal identification and damage assessment of a structure. In addition, reconstruction of more sparse de-noised signals is also studied and the proposed BCS-IPE algorithm is found to work even better for these more sparse signals.

**Acknowledgements**

This work was supported by the U.S. National Science Foundation under award number EAR-0941374 to the California Institute of Technology. This support is gratefully acknowledged by the first two authors. This research is also supported by grant from the National Natural Science Foundation of China (NSFC grant no. 51308161) and the International Postdoctoral Exchange Fellowship Program 2014 by the Office of China Postdoctoral Council, which partially supported the first author and this support is also gratefully acknowledged.

**Appendix A: Efficient updating formulae for hyper-parameter $\alpha_n$ for Algorithms BCS-MPE**

The following theory is based on [21] but for the prior PDF in (9) instead of (8).

In step 10 of Algorithm BCS-MPE, it is required to update $\boldsymbol{\mu}$ and $\boldsymbol{\Sigma}$, and $s_m$ and $q_m$, for all terms $(m = 1, ..., N)$. In the updating formulae, $s_m$ and $q_m$ are computed from:

$$s_m = S_m/(1 - \alpha_m^{-1} S_m), \tag{45}$$

$$q_m = Q_m/(1 - \alpha_m^{-1} S_m). \tag{46}$$

where the quantities $S_m$ and $Q_m$ are defined as:

$$S_m = \boldsymbol{\Theta}_m^T \mathbf{B}^{-1} \boldsymbol{\Theta}_m = \boldsymbol{\Theta}_m^T \boldsymbol{\Theta}_m - \beta \boldsymbol{\Theta}_m^T \boldsymbol{\Theta} \boldsymbol{\Sigma} \boldsymbol{\Theta}^T \boldsymbol{\Theta}_{nm} \tag{47}$$

$$Q_m = \boldsymbol{\Theta}_m^T \mathbf{B}^{-1} \mathbf{y} = \boldsymbol{\Theta}_n^T \mathbf{y} - \beta \boldsymbol{\Theta}_m^T \boldsymbol{\Theta} \boldsymbol{\Sigma} \boldsymbol{\Theta}^T \mathbf{y} \tag{48}$$

Efficient updating formulae for each potential implementation in steps 6-8 are given here and updated quantities are denoted by a hat (e.g. $\hat{\alpha}_n$).

(1) If $\hat{\gamma}_n > 0$ and $\alpha_n = \infty$, add $\boldsymbol{\Theta}_n$ in the model and update the corresponding $\alpha_n$, and:

$$\Delta \mathcal{L} = \frac{\beta Q_n^2 - S_n}{2S_n} + \frac{1}{2} \log \frac{S_n}{\beta Q_n^2}; \tag{49}$$

$$\hat{\boldsymbol{\Sigma}} = \begin{bmatrix} \boldsymbol{\Sigma} + \beta^2 \Sigma_{nn} \boldsymbol{\Sigma} \boldsymbol{\Theta}^T \boldsymbol{\Theta}_n \boldsymbol{\Theta}_n^T \boldsymbol{\Theta} \boldsymbol{\Sigma} & -\beta^2 \Sigma_{nn} \boldsymbol{\Sigma} \boldsymbol{\Theta}^T \boldsymbol{\Theta}_n \\ -\beta^2 \Sigma_{nn} (\boldsymbol{\Sigma} \boldsymbol{\Theta}^T \boldsymbol{\Theta}_n)^T & \Sigma_{nn} \end{bmatrix} \tag{50}$$

$$\hat{\boldsymbol{\mu}} = \begin{bmatrix} \boldsymbol{\mu} - \beta \mu_n \boldsymbol{\Sigma} \boldsymbol{\Theta}^T \boldsymbol{\Theta}_n \\ \mu_n \end{bmatrix} \tag{51}$$

where $\Sigma_{nn} = \beta^{-1}(\alpha_n + S_n)^{-1}, \mu_n = \beta \Sigma_{nn} Q_n;$

For each term $m = 1, ..., N$:

$$\hat{S}_m = S_m - \beta \Sigma_{nn} (\boldsymbol{\Theta}_m^T \mathbf{e}_n)^2 \tag{52}$$

$$\hat{Q}_m = Q_m - \mu_n (\boldsymbol{\Theta}_m^T \mathbf{e}_n) \tag{53}$$

where we define $\mathbf{e}_n = \boldsymbol{\Theta}_n - \beta \boldsymbol{\Theta} \boldsymbol{\Sigma} \boldsymbol{\Theta}^T \boldsymbol{\Theta}_n.$

(2) If $\hat{\gamma}_n > 0$ and $\alpha_n < \infty$, retain $\boldsymbol{\Theta}_n$ in the model and update the corresponding $\alpha_n$, and:

$$\Delta \mathcal{L} = \frac{\beta Q_n^2}{2\left(S_n + (\hat{\alpha}_n^{-1} - \alpha_n^{-1})^{-1}\right)} - \frac{1}{2} \log[1 + S_n(\hat{\alpha}_n^{-1} - \alpha_n^{-1})]; \tag{54}$$

$$\widehat{\boldsymbol{\Sigma}} = \boldsymbol{\Sigma} - \vartheta_j \boldsymbol{\Sigma}_j \boldsymbol{\Sigma}_j^T \tag{55}$$

$$\widehat{\boldsymbol{\mu}} = \boldsymbol{\mu} - \vartheta_j \mu_j \boldsymbol{\Sigma}_j \tag{56}$$

where we define $\vartheta_j = \left(\Sigma_{jj} + \beta^{-1}(\hat{\alpha}_n - \alpha_n)^{-1}\right)^{-1}$, $j$ denotes the index within the current basis (a smaller fraction of the full set of basis) that corresponds to the single term $n$ to be updated, and $\boldsymbol{\Sigma}_j$ is the $j$th column of $\boldsymbol{\Sigma}$.

For each term $m = 1, \dots, N$:

$$\hat{S}_m = S_m + \beta \vartheta_j \left(\boldsymbol{\Sigma}_j^T \boldsymbol{\Theta}^T \boldsymbol{\Theta}_m\right)^2 \tag{57}$$

$$\hat{Q}_m = Q_m + \vartheta_j \mu_j \left(\boldsymbol{\Sigma}_j^T \boldsymbol{\Theta}^T \boldsymbol{\Theta}_m\right) \tag{58}$$

(3) If $\hat{\gamma}_n \leq 0$ and $\alpha_n < \infty$, delete $\boldsymbol{\Theta}_n$ from the model and update the corresponding $\alpha_n$, and:

$$\Delta \mathcal{L} = \frac{\beta Q_n^2}{2(S_n - \alpha_n)} - \frac{1}{2} \log(1 - S_n \alpha_n^{-1}). \tag{59}$$

$$\widehat{\boldsymbol{\Sigma}} = \boldsymbol{\Sigma} - \frac{1}{\Sigma_{jj}} \boldsymbol{\Sigma}_j \boldsymbol{\Sigma}_j^T \tag{60}$$

$$\widehat{\boldsymbol{\mu}} = \boldsymbol{\mu} - \frac{\mu_j}{\Sigma_{jj}} \boldsymbol{\Sigma}_j \tag{61}$$

Following updates (60) and (61), the appropriate row and column are removed from $\widehat{\boldsymbol{\Sigma}}$ and $\widehat{\boldsymbol{\mu}}$.

For each term $m = 1, \dots, N$:

$$\hat{S}_m = S_m + \frac{\beta}{\Sigma_{jj}} \left(\boldsymbol{\Sigma}_j^T \boldsymbol{\Theta}^T \boldsymbol{\Theta}_m\right)^2 \tag{62}$$

$$\hat{Q}_m = Q_m + \frac{\mu_j}{\Sigma_{jj}} \left(\boldsymbol{\Sigma}_j^T \boldsymbol{\Theta}^T \boldsymbol{\Theta}_m\right) \tag{63}$$

**Appendix B: Efficient updating formulae for hyper-parameter $\alpha_n$ for Algorithms BCS-IPE**

In the implementation for updating hyper-parameter $\alpha_n$ for Algorithm BCS-IPE, the quantities $\boldsymbol{\Lambda}$, $q_m$, $s_m$, and $g_m$ for all terms $(m = 1, \dots, N)$ are needed in each iteration. Efficient updating formulae are presented here following [26]. Updated quantities are denoted by a tilda (e.g. $\tilde{\alpha}_n$).

In the updating formulae, $g_m$ is computed from:

$$g_m = G_m + Q_m^2 / (\alpha_m - S_m) \tag{64}$$

$$G_m = \mathbf{y}^T \mathbf{B}^{-1} \mathbf{y} + 2b_0 = \mathbf{y}^T \mathbf{y} - \mathbf{y}^T \mathbf{\Theta \Lambda \Theta}^T \mathbf{y} + 2b_0 \tag{65}$$

The other quantities $S_m, Q_m, s_m,$ and $q_m$ can be found in (45-48).

(1) If $\tilde{\gamma}_n > 0$ and $\alpha_n = \infty$, add vector $\mathbf{\Theta}_n$ and update $\alpha_n$, where:

$$\Delta \mathcal{L} = \frac{1}{2} \log \frac{\alpha_n}{\alpha_n + s_n} - \frac{1}{2}(K + 2a_0) \log\left(1 - \frac{q_n^2/g_n}{\alpha_n + s_n}\right); \tag{66}$$

$$\tilde{\mathbf{\Lambda}} = \begin{bmatrix} \mathbf{\Lambda} + \Lambda_{nn}\mathbf{\Lambda \Theta}^T \mathbf{\Theta}_n \mathbf{\Theta}_n^T \mathbf{\Theta \Lambda} & -\Lambda_{nn}\mathbf{\Lambda \Theta}^T \mathbf{\Theta}_n \\ -\Lambda_{nn}(\mathbf{\Lambda \Theta}^T \mathbf{\Theta}_n)^T & \Lambda_{nn} \end{bmatrix} \tag{67}$$

$$\tilde{\boldsymbol{\mu}} = \begin{bmatrix} \boldsymbol{\mu} - \mu_n \mathbf{\Lambda \Theta}^T \mathbf{\Theta}_n \\ \mu_n \end{bmatrix} \tag{68}$$

where $\Lambda_{nn} = (\alpha_n + S_n)^{-1}, \mu_n = \Lambda_{nn} Q_n;$

For each term $m = 1, \ldots, N$:

$$\tilde{S}_m = S_m - \Lambda_{nn}(\mathbf{\Theta}_m^T \boldsymbol{\xi}_n)^2 \tag{69}$$

$$\tilde{Q}_m = Q_m - \mu_n(\mathbf{\Theta}_m^T \boldsymbol{\xi}_n) \tag{70}$$

$$\tilde{G}_m = G_m - \Lambda_{nn}(\mathbf{y}^T \boldsymbol{\xi}_n)^2 \tag{71}$$

where we define $\boldsymbol{\xi}_n = \mathbf{\Theta}_n - \mathbf{\Theta \Lambda \Theta}^T \mathbf{\Theta}_n$.

(2) If $\tilde{\gamma}_n > 0$ and $\alpha_n < \infty$, update $\alpha_n$:

$$\Delta \mathcal{L} = \frac{1}{2}(K + 2a_0 - 1)\log[1 + S_n(\tilde{\alpha}_n^{-1} - \alpha_n^{-1})] + \frac{1}{2}(K + 2a_0) \log \frac{[(\alpha_n + s_n)g_n - q_n^2]\tilde{\alpha}_n}{[(\tilde{\alpha}_n + s_n)g_n - q_n^2]\alpha_n} \tag{72}$$

$$\tilde{\mathbf{\Lambda}} = \mathbf{\Lambda} - \gamma_j \mathbf{\Lambda}_j \mathbf{\Lambda}_j^T \tag{73}$$

$$\tilde{\boldsymbol{\mu}} = \boldsymbol{\mu} - \gamma_j \mu_j \mathbf{\Lambda}_j \tag{74}$$

where we define $\gamma_j = \left(\Lambda_{jj} + (\tilde{\alpha}_n - \alpha_n)^{-1}\right)^{-1}$, $j$ denotes the index within the current basis that corresponds to the single term $n$ to be updated, and $\mathbf{\Lambda}_j$ is the *jth* column of $\mathbf{\Lambda}$.

For each term $m = 1, \ldots, N$:

$$\tilde{S}_m = S_m + \gamma_j \left(\mathbf{\Lambda}_j^T \mathbf{\Theta}^T \mathbf{\Theta}_m\right)^2 \tag{75}$$

$$\tilde{Q}_m = Q_m + \gamma_j \mu_j \left(\mathbf{\Lambda}_j^T \mathbf{\Theta}^T \mathbf{\Theta}_m\right) \tag{76}$$

$$\tilde{G}_m = G_m + \gamma_j\left(\Lambda_j^T \Theta^T \Theta_m\right)^2 \tag{77}$$

(3) If $\tilde{\gamma}_n \leq 0$ and $\alpha_n < \infty$, delete vector $\Theta_n$:

$$\Delta\mathcal{L} = -\frac{1}{2}(K + 2a_0)\log\left(1 + \frac{Q_n^2/G_n}{\alpha_n - S_n}\right) - \frac{1}{2}\log(1 - S_n\alpha_n^{-1}) \tag{78}$$

$$\tilde{\Lambda} = \Lambda - \frac{1}{\Lambda_{jj}}\Lambda_j\Lambda_j^T \tag{79}$$

$$\tilde{\mu} = \mu - \frac{\mu_j}{\Lambda_{jj}}\Lambda_j \tag{80}$$

Following updates (79) and (80), the appropriate row and column are removed from $\tilde{\Lambda}$ and $\tilde{\mu}$.

For each term $m = 1, \ldots, N$:

$$\tilde{S}_m = S_m + \frac{1}{\Lambda_{jj}}\left(\Lambda_j^T \Theta^T \Theta_m\right)^2 \tag{81}$$

$$\tilde{Q}_m = Q_m + \frac{\mu_j}{\Lambda_{jj}}\left(\Lambda_j^T \Theta^T \Theta_m\right) \tag{82}$$

$$\tilde{G}_m = G_m + \frac{1}{\Lambda_{jj}}\left(\Lambda_j^T \Theta^T \mathbf{y}\right)^2 \tag{83}$$